\documentclass[
 reprint,
 amsmath,amssymb,
 aps,
 prd,
]{revtex4-2}

\usepackage{dcolumn}
\usepackage{indentfirst}
\usepackage{amsmath}
\usepackage{bm}
\usepackage{slashed}
\usepackage{graphicx}
\usepackage{hyperref}

\begin{document}

\preprint{APS/123-QED}

\title{Alleviating Cosmological Tensions with a Coupled Scalar Fields Model}

\author{Gang Liu}
 \email{liugang\_dlut@mail.dlut.edu.cn}
\author{Zhihuan Zhou}%
\author{Yuhao Mu}%

\author{Lixin Xu}
 \email{lxxu@dlut.edu.cn}
\affiliation{%
 Institute of Theoretical Physics\\
 School of Physics\\
 Dalian University of Technology\\
 Dalian 116024, People's Republic of China
}

\date{\today}

\begin{abstract}
In this paper, we investigate the interaction between early dark energy (EDE) and scalar 
field dark matter, proposing a coupled scalar fields model to
address the Hubble tension and $S_8$ tension. While the EDE model successfully alleviates 
the Hubble tension, it exacerbates 
the $S_8$ tension. To mitigate the negative impact of EDE, 
we introduce the interaction between EDE and dark matter. Specifically, we replace cold 
dark matter with scalar field dark matter, given its capability to suppress structure 
growth on small scales.
We constrained the new model using cosmological observations including 
the temperature and polarization anisotropy power spectra data of cosmic microwave background
radiation (CMB) from \textit{Planck} 2018 results, 
baryon acoustic oscillations (BAO) measurements extracted from 6dFGS, SDSS and BOSS, 
the Pantheon sample of type Ia supernovae (SNIa), 
the local distance-ladder data (SH0ES), 
and the Dark Energy Survey Year-3 data. 
Employing Markov Chain Monte Carlo method, we find that this novel model yields 
best-fit values of $H_0$ and $S_8$ equal to $71.13$ km/s/Mpc and $0.8256$, respectively.
Compared to the $\Lambda$CDM model, the new model alleviates the Hubble tension but 
still fails to resolve the $S_8$ tension. However, we obtain a smaller value of $S_8$ 
compared to the result of $0.8316$ obtained for EDE model, which 
mitigates to some extent the shortcoming of the EDE model.
\end{abstract}

\maketitle


\section{Introduction}
Over the past few years, the $\Lambda$CDM model has encountered numerous challenges as a 
result of the growing quantity and quality of observations. 
The emergence of the Hubble tension and the $S_8$ tension has garnered significant attention. 
The Hubble tension \cite{Verde2019TensionsBT} pertains to the discrepancy between the 
$H_0$ value obtained from model-independent local measurements such as Type Ia supernovae 
(SNIa) \cite{Riess_2019, Riess_2021}, and the $H_0$ value derived from the cosmic 
microwave background (CMB) \cite{planck2020} and the large-scale structure (LSS) 
\cite{Cooke_2016, Sch_neberg_2019, PRD.92.123516, Philcox_2020}.
More precisely, the \textit{Planck} 2018 CMB data estimates the value of $H_0$ to be 
$67.37\pm0.54$ km/s/Mpc \cite{planck2020}, while the cosmic distance ladder measurement 
(SH0ES) yields $H_0=73.04\pm1.04$ km/s/Mpc \cite{Riess2021ACM}, with a statistical error 
of 4.8$\sigma$. 

The $S_8$ tension characterizes the inconsistency between CMB and LSS observations 
\cite{PRL.111.161301, Hildebrandt_2020}. 
The \textit{Planck} best-fit $\Lambda$CDM model estimates the value of $S_8$ to be 
$0.834\pm0.016$ \cite{planck2020}, while LSS observations yield 
$0.759^{+0.024}_{-0.021}$ for KiDS-1000 \cite{KiDS}, $0.800^{+0.029}_{-0.028}$ for 
HSC-Y1 \cite{psz010}, and $0.776\pm0.017$ for Dark Energy Survey Year-3 (DES-Y3) 
\cite{PRD.105.023520}. 

Numerous models have been proposed to address the Hubble tension, as reviewed recently 
by \cite{Di_2021}. These models incorporate modifications to the late universe, such as 
the Phenomenologically Emergent Dark Energy model \cite{Li_2019}, the Phantom Transition 
\cite{Zhouzh}, and the early universe, including the Early Dark Energy (EDE) model 
\cite{PRL.122.221301, PRD.101.063523, PRD.102.043507}, the Acoustic Dark Energy model 
\cite{PRD.100.063542}, and the New Early Dark Energy model \cite{PhysRevD.103.L041303}, etc.

Despite the proposed models, they still encounter several issues. For instance, the 
late-time solutions that do not alter the sound horizon are 
generally unable to account for the SH0ES measurement. Conversely, the early-time 
solutions that introduce a new component before recombination to decrease the scale 
of the acoustic horizon on the final scattering surface, increase the value of $H_0$, 
and maintain the angular scale of the acoustic horizon in CMB observations, but they 
exacerbate the $S_8$ tension \cite{PRD.102.043507}. 

\begin{figure}
    \includegraphics[width=\columnwidth]{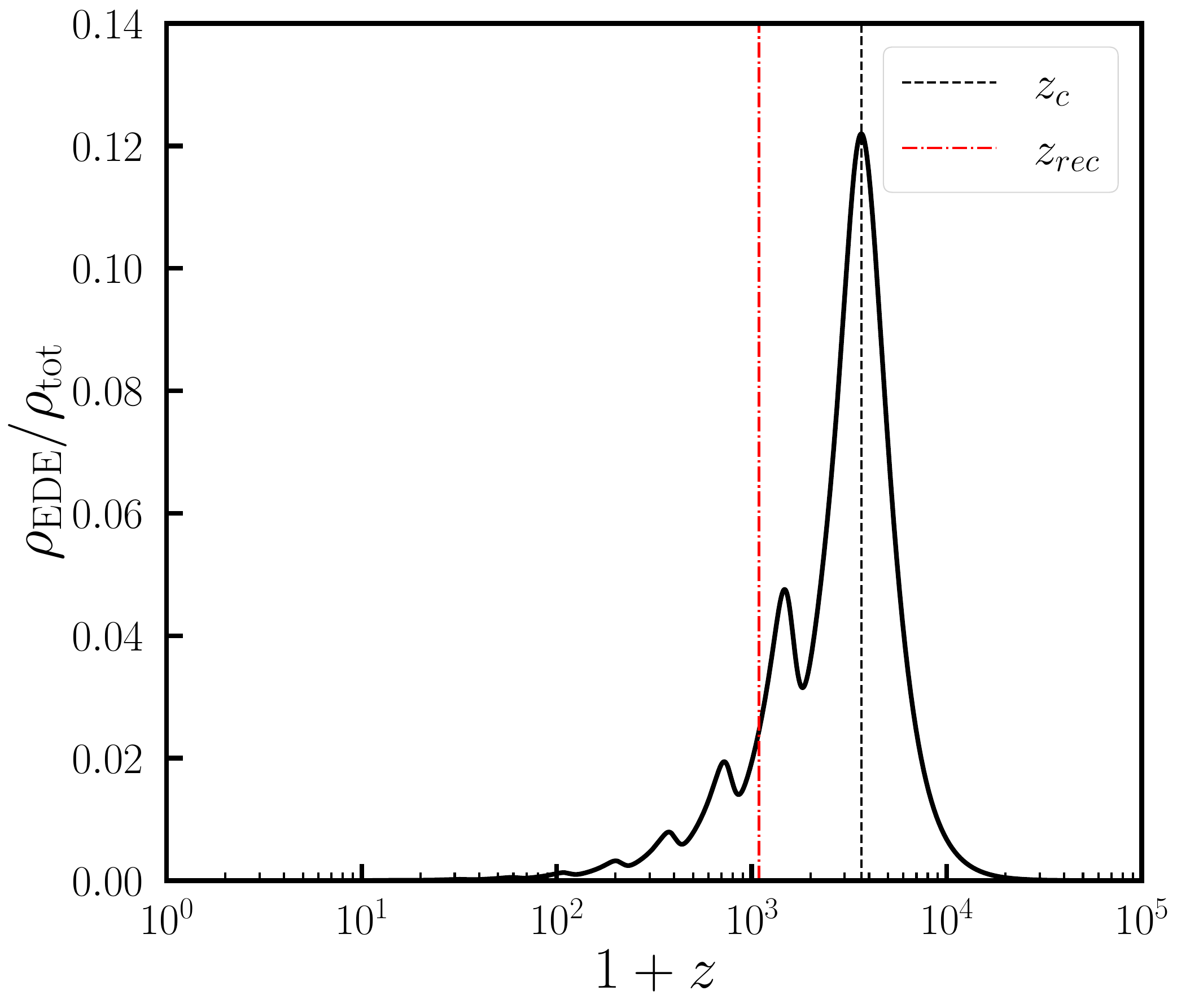}
    \caption{\label{fig1}The evolution of the EDE energy density 
    as a fraction of the total energy density with respect to the redshift. The red vertical 
    dash-dotted line corresponds to the redshift at recombination $z_\mathrm{rec}$, while the 
    black vertical dashed line represents the critical redshift of EDE.}
\end{figure}

This paper focuses on the EDE model and aims to address the associated concerns. 
EDE is characterized by an ultra-light axion scalar field \cite{PRL.113.251302, MARSH20161}. 
In this model, $z_c$ denotes the redshift at the apex of the EDE component contribution, 
while $f_\mathrm{EDE}$ represents the proportion of EDE energy density relative to 
the total energy density at that time. The evolution of the EDE fraction with the redshift 
is depicted in Fig.~\ref{fig1}, where the red vertical dashed line denotes the redshift 
at recombination, and the apex of the EDE component occurs before recombination. The 
Hubble tension can be resolved when the EDE ratio $f_\mathrm{EDE}$ reaches approximately 
10\% \cite{PRD.102.043507}.

During its contribution period, the EDE component marginally diminishes the perturbed 
growth of the structure. To align with the CMB data, the cold dark matter (CDM) density 
must be augmented to offset these losses. Moreover, some other cosmological parameters, 
such as the scalar spectral index $n_s$, the baryon density 
$\omega_b$, and the amplitude of density perturbations $\sigma_8$, will also undergo 
changes \cite{2112.09128}. Consequently, the EDE model will invariably exacerbate the 
CMB-LSS inconsistency \cite{PRD.102.043507, D_Amico_2021}. 

To reduce the $S_8$ tension, it is common to investigate the interaction of dark matter 
(DM) and dark energy (DE), which can inhibit the growth of structure through the drag of 
DE on DM \cite{PRD.104.083510, PRD.104.103503}. In addition, since the nature of dark 
matter is not yet comprehensively understood, alternative descriptions can be developed 
to substitute cold dark matter, and alleviate the $S_8$ tension.

The $\Lambda$CDM model posits that dark matter is comprised of non-baryonic, pressureless, 
and non-relativistic particles \cite{2112.09337}. This model has been successful in 
explaining large-scale observations from the cosmic microwave background (CMB) and 
large-scale structure (LSS). However, despite its achievements, the microscopic 
properties of dark matter remain unknown \cite{2005.03254}. The aforementioned 
assumptions have led to a number of unresolved issues, such as the unexpected behavior 
of central densities in galactic halos and the overpopulation of secondary structures 
on small scales. These observations suggest that the cold dark matter (CDM) may not be an 
adequate description of dark matter, particularly on smaller scales \cite{2112.09337}.

Scalar field dark matter (SFDM) presents an alternative to CDM, which is composed of a 
light scalar field with a mass of approximately $10^{-22}$ eV 
\cite{2005.03254, 2112.09337}. In this model, the scalar field forms a Bose-Einstein 
condensate (BEC) at the galactic scale, which modifies the dynamics of dark matter on 
small scales while maintaining the success of CDM on large scales. This condensation 
leads to the suppression of structure growth on small scales, which could potentially 
alleviate the $S_8$ tension.

\begin{figure}
    \includegraphics[width=\columnwidth]{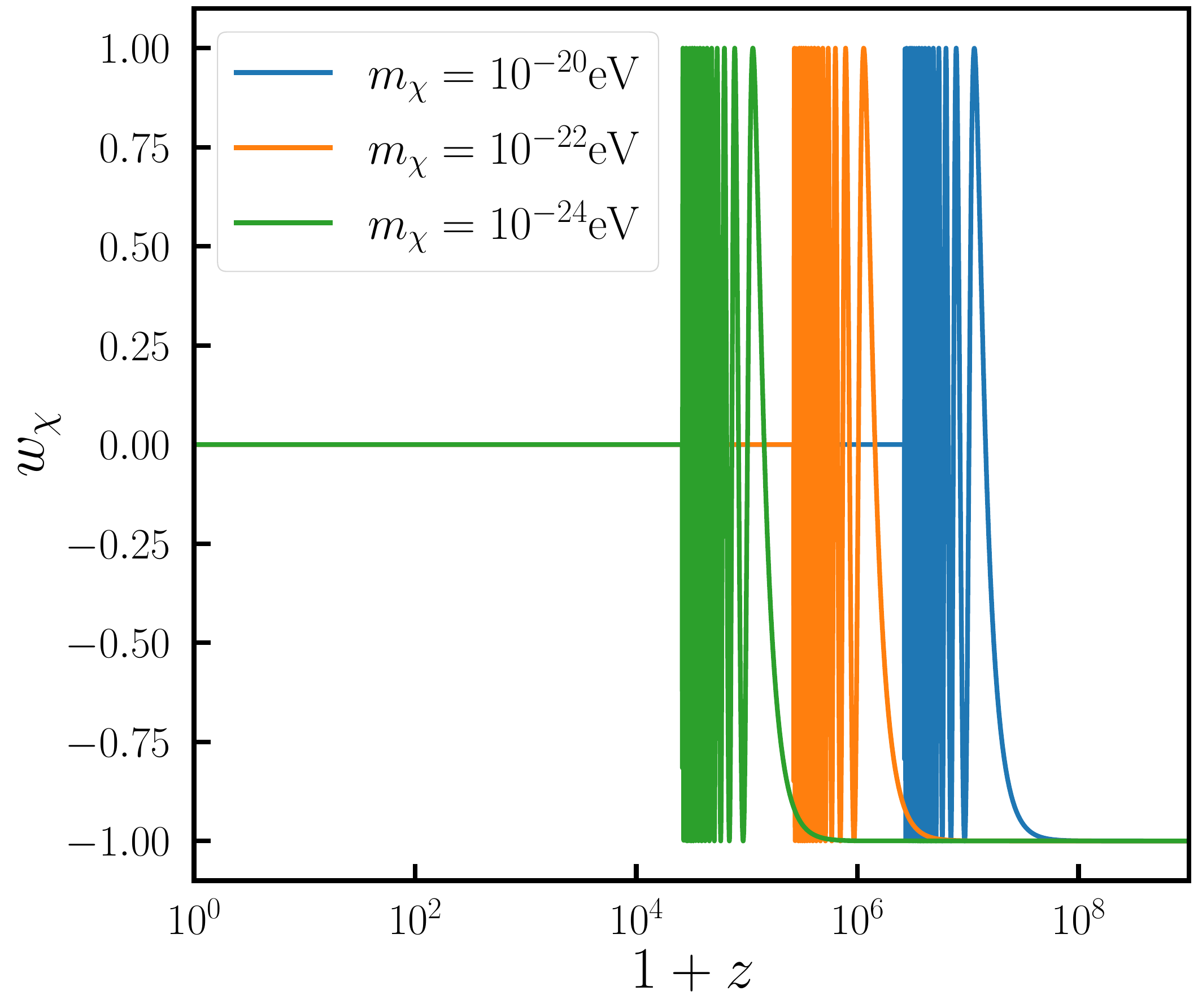}
    \caption{\label{fig2}The equation of state of scalar field dark matter is 
    presented as a function of the redshift for various masses. 
    Initially, SFDM behaves like the cosmological constant, followed 
    by oscillations, and ultimately evolves into a behavior similar to cold dark 
    matter. The initiation time of the oscillations is dependent on the field mass.}
\end{figure}

The behavior of the scalar field dark matter is similar to that of the cosmological 
constant in the early universe, followed by oscillations, and ultimately similar to 
CDM. Fig.~\ref{fig2} displays the evolution of the SFDM equation of 
state with the redshift. The initiation time of the oscillations is determined 
by the field mass. A smaller mass results in later oscillations.

This paper proposes a coupled scalar fields (CSF) model that explores the interaction 
between early dark energy (EDE) and scalar field dark matter (SFDM). The coupling 
between the two fields is inspired by the Swampland Distance Conjecture (SDC) 
\cite{Vafa2005TheSL,Palti}, which has previously been applied to quintessence 
models \cite{PRD.103.043523} and the EDE model \cite{2112.09128}. 
According to the SDC, a low-energy effective field theory is deemed valid only within 
a region of field space constrained by the Planck scale. Moreover, any breakdown of 
the effective field theory that arises due to Planckian field excursions can be 
expressed as an exponential sensitivity reflected in the mass spectrum of the 
effective theory. Specifically, 
the CSF model posits that the mass of dark matter is exponentially dependent on 
the EDE scalar:
\begin{equation} \label{eq1}
    m_\mathrm{DM}(\phi)=m_0e^{\beta\phi/M_{pl}},
\end{equation}
where, $m_0$ represents the present-day mass of dark matter, $\phi$ denotes the EDE 
scalar, $\beta \sim \mathcal{O}(1)$ is a constant, and $M_{pl}=2.435\times10^{27}$ eV 
denotes the reduced Planck mass.

In this study, we have conducted a comprehensive investigation into the evolutionary 
equations of the coupled model at both the background and perturbation levels. We employed 
a Markov Chain Monte Carlo (MCMC) analysis of three 
cosmological models, namely the $\Lambda$CDM, EDE, and CSF models. We utilized 
various datasets, including the \textit{Planck} 2018 primary CMB data and CMB lensing 
data \cite{planck2020,osti_1676388,osti_1775409}, BAO measurements from the BOSS DR12, 
the 6dF galaxy survey, and SDSS DR7 \cite{Alam_2017,Buen_Abad_2018,19250.x,10.1093}, 
the Pantheon supernovae Ia data \cite{Scolnic_2018}, the SH0ES measurement 
\cite{Riess2021ACM}, and the Dark Energy Survey Year-3 data \cite{PRD.105.023520}.

Based on the entire datasets, we found that the $H_0$ values obtained by the EDE and 
CSF models are $72.46\pm 0.86$ km/s/Mpc and $72.20\pm 0.81$ km/s/Mpc at a 68\% 
C.L., respectively, both exceeding the result of $68.71^{+0.35}_{-0.41}$ km/s/Mpc 
obtained by the $\Lambda$CDM model. Therefore, both models can alleviate the Hubble 
tension. The $S_8$ value for the EDE model is $0.822^{+0.011}_{-0.0093}$, while the 
result for CSF is $0.820^{+0.014}_{-0.008}$. Furthermore, the obtained coupling 
constant is constrained to be $-0.014\pm 0.016$, indicating an interaction between 
dark matter and dark energy.
Despite the failure of the coupled model to resolve the $S_8$ 
tension, it has yielded a smaller $S_8$ and $\chi^2_\mathrm{tot}$ compared to the EDE 
model, thereby mitigating the adverse effect associated with EDE.

The paper is organized as follows. Section \ref{sec2} presents an introduction to 
the CSF model, including the dynamics of background and perturbation. 
In Section \ref{sec3}, we present numerical results illustrating the impact of the 
coupled model on the large-scale structures. 
In Section \ref{sec4}, we discuss the datasets utilized in our analysis and present the 
corresponding results. Finally, we summarize our findings in Section 
\ref{sec5}.

\section{Coupled Scalar Fields}\label{sec2}
We examine the coupling between SFDM and EDE. The Lagrangian 
is defined as follows:
\begin{equation}
    \mathcal{L}=-\frac{1}{2}\partial^{\mu}\chi\partial_{\mu}\chi-\frac{1}{2}m_{\chi}(\phi)^2\chi^2-\frac{1}{2}\partial^{\mu}\phi\partial_{\mu}\phi-V(\phi),    
\end{equation}
where $\phi$ is the EDE scalar with the potential \cite{PRD.102.043507}
\begin{equation}
    V(\phi)=m_{\phi}^2f_{\phi}^2[1-\cos(\phi/f_{\phi})]^3+V_{\Lambda}, \label{Vp}
\end{equation}
and $\chi$ is SFDM scalar with a $\phi$-dependent mass $m_{\chi}(\phi)$, $V_{\Lambda}$ 
in Eq.(\ref{Vp}) serves as the cosmological constant. 
The subscript $\phi$ is used to denote dark energy and $\chi$ is used to denote dark matter. 
Numerous potentials of SFDM have been investigated in previous studies 
\cite{BEYER2014418,AMENDOLA2006192}, but the common features of them can be represented 
by $\frac{1}{2}m_{\chi}^2\chi^2$ \cite{Ure_a_L_pez_2016}. The specific form of $m_{\chi}(\phi)$ is 
\begin{equation}
    m_{\chi}(\phi)=m_0e^{\beta\phi/M_{pl}},
\end{equation}
which is given by the Swampland Distance Conjecture as Eq.(\ref{eq1}), and $m_0$ 
represents the present-day mass of SFDM. 

\subsection{Background Equations}
The motion equations of the scalar field dark matter (SFDM) in a flat 
Friedmann-Lema\^itre-Robertson-Walker (FLRW) cosmology can be expressed as follows:
\begin{subequations}
    \begin{align}
        &3M_{pl}^2H^2=\sum_I\rho_I, \label{fe1}\\
        &-2M_{pl}^2\dot{H}=\sum_I{\rho_I+p_I}, \label{fe2}\\
        &\ddot{\chi}=-3H\dot{\chi}-m_{\chi}^2\chi,\label{KG}
    \end{align} 
\end{subequations}
where the dot denotes the derivative with respect to cosmic time, and $H$ is the
Hubble parameter, $\rho_I$ and $p_I$ are the energy density and pressure for each 
component respectively.
The expressions for the energy density and pressure of SFDM are as follows:
\begin{subequations}\label{rhop}
    \begin{align}
        &\rho_{\chi}=\frac{1}{2}\dot{\chi}^2+\frac{1}{2}m_{\chi}^2\chi^2,\\
        &p_{\chi}=\frac{1}{2}\dot{\chi}^2-\frac{1}{2}m_{\chi}^2\chi^2.
    \end{align} 
\end{subequations}

We define a new set of variables to transform the Klein-Gordon equation (\ref{KG}) 
\cite{PRD.57.4686},
\begin{equation}
    x=\frac{\dot{\chi}}{\sqrt{6}M_{pl}H}, 
    \quad y=-\frac{m_{\chi}\chi}{\sqrt{6}M_{pl}H}, 
    \quad y_1=\frac{2m_{\chi}}{H}.
\end{equation}
We utilize the polar coordinate variable transformation form as proposed in previous 
works \cite{Ure_a_L_pez_2016,PRD.57.4686,PRD.94.063532}:
\begin{equation}
    x=\sqrt{\Omega_{\chi}}\sin{\frac{\theta}{2}}, \quad
    y=\sqrt{\Omega_{\chi}}\cos{\frac{\theta}{2}}, \quad
\end{equation}
where $\Omega_{\chi}=\frac{\rho_{\chi}}{3M_{pl}^2H^2}$ is the density parameter 
of the dark matter.
The Friedman equations (\ref{fe1}) and (\ref{fe2}) are reformulated as follows:
\begin{equation}
    \frac{\dot{H}}{H^2}=-\frac{3}{2}(1+w_t), \quad
    1=\sum_I\Omega_I+\Omega_{\chi},
\end{equation}
where 
$w_t=\frac{p_t}{\rho_t}$ represents the total equation of state, 
which is the ratio of total pressure $p_t$ to total energy density $\rho_t$, 
and $\Omega_I=\frac{\rho_I}{3M_{pl}^2H^2}$ 
is the density parameter of each components.
The Klein-Gordon equation (\ref{KG}) becomes:
\begin{subequations}
    \begin{align}
        &\frac{\dot{\Omega}_{\chi}}{\Omega_{\chi}}=3H(w_t+\cos\theta)+\frac{\beta\dot{\phi}}{M_{pl}}(1+\cos\theta),\\
        &\dot{\theta}=H(y_1-3\sin\theta)-\frac{\beta\dot{\phi}}{M_{pl}}\sin\theta,\\
        &\dot{y_1}=\frac{3}{2}H(1+w_t)y_1+\frac{\beta\dot{\phi}}{M_{pl}}y_1.
    \end{align} 
\end{subequations}

The equations of motion for the EDE is given by the variation of 
the action expanded to linear order in $\delta \phi$, 
\begin{equation}
    \ddot{\phi}+3H\dot{\phi}+\frac{dV}{d\phi}=-3\beta M_{pl}H^2\Omega_{\chi}(1+\cos\theta) \label{KGDE}.
\end{equation}

\begin{figure*}
    \includegraphics[width=\linewidth]{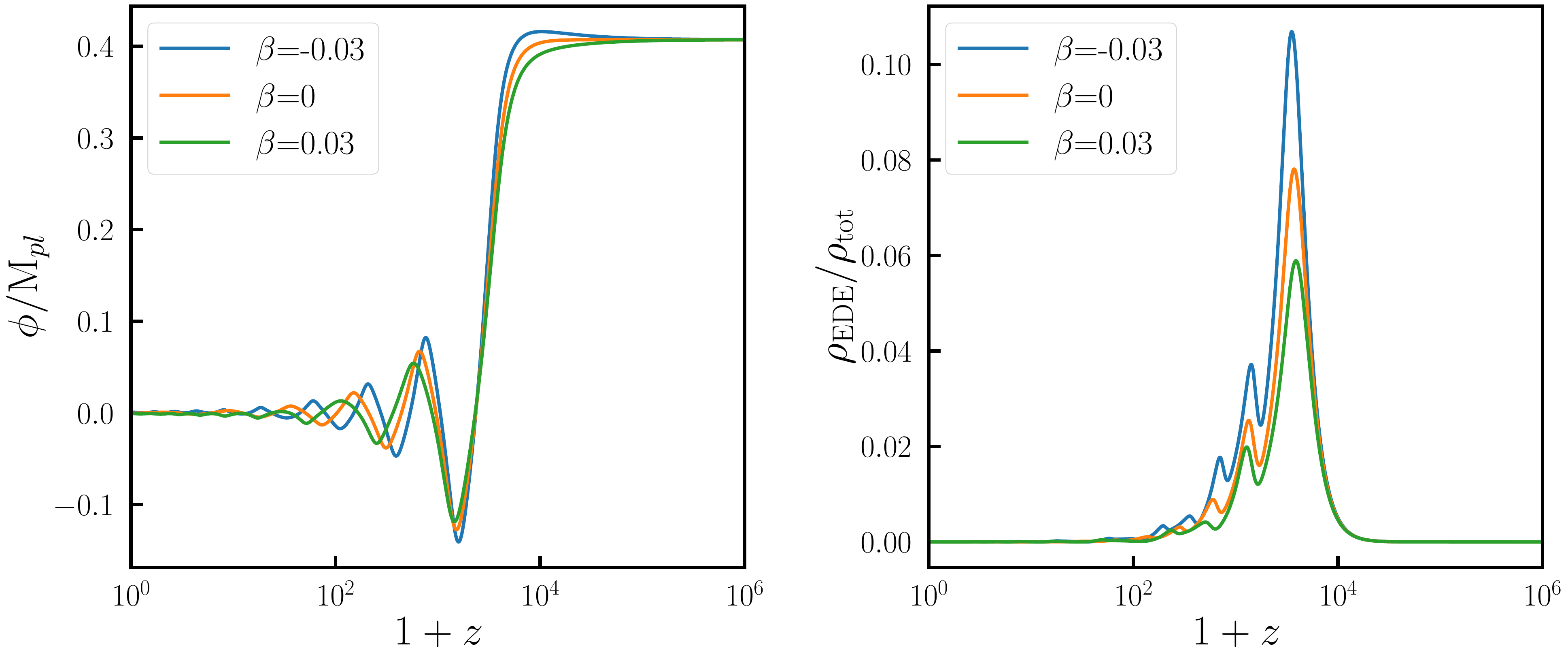}
    \caption{\label{fig3}The variation of the EDE scalar 
    (in the left panel) and the EDE energy density fraction 
    (in the right panel) as functions of the redshift, for different coupling 
    constants. The negative (positive) coupling constant leads to the conversion 
    of the dark matter (dark energy) component into dark energy (dark matter), 
    increasing (decreasing) the EDE energy density fraction. }
\end{figure*}

The left panel of Fig.~\ref{fig3} depicts the evolution of the EDE scalar, 
while the right panel shows the EDE energy density fraction as a function of the 
redshift across various coupling constants. The cosmological parameters utilized in this 
analysis are derived from the best-fit values listed in Tab.~\ref{tab1}. 
The amplitude and phase of the EDE scalar will be altered by varying coupling constants. 
The sign of the coupling constant determines the direction of conversion between dark 
matter and dark energy components. A negative coupling constant results in a source 
term on the right-hand side of Eq.(\ref{KGDE}), causing the conversion of dark matter 
into dark energy and leading to an increase in the energy density fraction of EDE. 
Conversely, a positive coupling constant causes the conversion of dark energy into 
dark matter.

The energy density and pressure of the EDE are,
\begin{subequations}
    \begin{align}
        &\rho_{\phi}=\frac{1}{2}\dot{\phi}^2+V(\phi),\\
        &p_{\phi}=\frac{1}{2}\dot{\phi}^2-V(\phi).
    \end{align} 
\end{subequations}
The equations of continuity for SFDM and EDE can be derived from the Klein-Gordon 
equations presented in Eq.(\ref{KG}) and Eq.(\ref{KGDE}), respectively,
\begin{subequations}
    \begin{align}
        &\dot{\rho_{\chi}}=-3H(\rho_{\chi}+p_{\chi})+\frac{\beta\dot{\phi}}{M_{pl}}(1+\cos\theta)\rho_{\chi},\\
        &\dot{\rho_{\phi}}=-3H(\rho_{\phi}+p_{\phi})-\frac{\beta\dot{\phi}}{M_{pl}}(1+\cos\theta)\rho_{\chi}.
    \end{align}
\end{subequations}
Many coupled dark energy models have this common form \cite{Wang_2016, Kazuya_Koyama_2009, 
PhysRevD.96.123508, Mukherjee2016InSO, PhysRevD.77.063513}, which ensures covariant 
conservation for the total stress tensor.

\subsection{Perturbution Equations}
We calculated the perturbation equation using the synchronous gauge, where the metric 
is defined as follows:
\begin{equation}
    ds^2=-dt^2+a^2(t)(\delta_{ij}+h_{ij})dx^idx^j.
\end{equation}
The scalars of SFDM and EDE are given by 
\begin{subequations}
    \begin{align}
        &\chi(x,t)=\chi(t)+\delta\chi(x,t),\\
        &\phi(x,t)=\phi(t)+\delta\phi(x,t),
    \end{align} 
\end{subequations}
with $\chi(t)$, $\phi(t)$ the background parts, and $\delta\chi(x,t)$, 
$\delta\phi(x,t)$ the linear perturbations respectively.

The perturbed Klein-Gordon equation for a Fourier mode of $\delta\chi(x,t)$ is
\begin{equation}\label{lkg}
    \ddot{\delta\chi}=-3H\dot{\delta\chi}-(\frac{k^2}{a^2}+m_{\chi}^2)\delta\chi-\frac{1}{2}\dot{\chi}\dot{h}.
\end{equation}
where $h$ representes the trace of scalar metric perturbations. 
The density perturbations $\delta\rho(\chi)$, pressure perturbations $\delta p(\chi)$, 
and velocity divergence $\Theta(\chi)$ can be expressed as provided in 
\cite{PRD.58.023503,Hu_1998},
\begin{subequations}\label{rpt}
    \begin{align}
        &\delta \rho_{\chi}=\dot{\chi}\dot{\delta \chi}+\partial_{\chi}V(\chi)\delta \chi,\\
        &\delta p_{\chi}=\dot{\chi}\dot{\delta \chi}-\partial_{\chi}V(\chi)\delta \chi,\\
        &(\rho_{\chi}+p_{\chi})\Theta_{\chi}=\frac{k^2}{a}\dot{\chi}\delta \chi.
    \end{align} 
\end{subequations}

As previously done in the background section, we introduce new variables to derive 
the perturbation equation \cite{PRD.94.063532}:
\begin{subequations}
    \begin{align}
        &u=\sqrt{\frac{2}{3}}\frac{\dot{\delta \chi}}{M_{pl}H}=-\sqrt{\Omega_{\chi}}e^{\alpha}\cos\frac{\vartheta}{2}\\
        &v=\sqrt{\frac{2}{3}}\frac{m_{\chi}\delta \chi}{M_{pl}H}=-\sqrt{\Omega_{\chi}}e^{\alpha}\sin\frac{\vartheta}{2}
    \end{align} 
\end{subequations}
Once more, we introduce a new set of variables:
\begin{subequations}
    \begin{align}
        &\delta_0=-e^{\alpha}\sin(\frac{\theta}{2}-\frac{\vartheta}{2}),\\
        &\delta_1=-e^{\alpha}\cos(\frac{\theta}{2}-\frac{\vartheta}{2}).
    \end{align} 
\end{subequations}
The equations of motion for the new variables are:
\begin{widetext}
    \begin{subequations}
        \begin{align}
            &\dot{\delta_{0}}=\delta_0H\omega\sin\theta-\delta_1[3H\sin\theta+H\omega(1-\cos\theta)]-\frac{\dot{h}}{2}(1-\cos\theta)-\frac{\beta\dot{\phi}}{M_{pl}}\delta_1\sin\theta \label{d0},\\
            &\dot{\delta_{1}}=\delta_0H\omega(1+\cos\theta)-\delta_1(3H\cos\theta+H\omega\sin\theta)-\frac{\dot{h}}{2}\sin\theta-\frac{\beta\dot{\phi}}{M_{pl}}\delta_1\cos\theta,
        \end{align} 
    \end{subequations}
\end{widetext}
where 
\begin{equation}
    \omega=\frac{k^2}{2a^2m_{\chi}H}=\frac{k^2}{a^2H^2y_1}.
\end{equation}
The relationship between the new variables and density, pressure, and velocity 
divergence can be established by referring to the definition given in Eq.(\ref{rpt}),
\begin{subequations}
    \begin{align}
        &\delta\rho_{\chi}=\rho_{\chi}\delta_0,\\
        &\delta p_{\chi}=\rho_{\chi}(\delta_1\sin\theta-\delta_0\cos\theta),\\
        &(\rho_{\chi}+p_{\chi})\Theta_{\chi}=\frac{k^2}{aHy_1}\rho_{\chi}[\delta_1(1-\cos\theta)-\delta_0\sin\theta].
    \end{align}
\end{subequations}

Expanding the action to the second order and varying with respect to $\delta \phi$, we 
obtain the equation of motion for EDE perturbation is,
\begin{widetext}
    \begin{equation}
        \ddot{\delta \phi}+3H\dot{\delta \phi}+\frac{1}{2}\dot{h}\dot{\phi}+(\frac{k^2}{a^2}+\frac{d^2V}{d\phi^2})\delta \phi
        =\frac{\beta}{M_{pl}}\rho_{\chi}[\delta_1\sin\theta-\delta_0(1+\cos\theta)]-2(\frac{\beta}{M_{pl}})^2\rho_{\chi}(1+\cos\theta)\delta \phi.
    \end{equation}
\end{widetext}

\subsection{Initial Conditions}
In the early universe, Hubble friction induces the effective freezing of the scalar 
fields at their initial value, leading to a slow-roll process. The 
initial value of $\dot{\phi}$ can be set to 0. And the energy 
density of dark matter can be approximated to 
be negligible at that time. 
As a result, the equations of the EDE and SFDM simplify to an uncoupled form. 

We treat the ratio between the initial value of EDE scalar and the axion decay constant, 
$\alpha_i=\phi_i/f_{\phi}$ as the model parameter \cite{PRD.101.063523,PRD.102.043507}. 
We employ the attractor solution initial conditions for SFDM, namely, 
\begin{subequations}
    \begin{align}
        &\theta_i=\frac{2}{5}\frac{m_0e^{\frac{\beta\phi_i}{M_{pl}}}}{H_0\sqrt{\Omega_r a_i^{-4}}},\\
        &y_{1i}=5\theta_i,
    \end{align}
\end{subequations}
where $\Omega_r$ represents the energy density fraction of the present radiation 
component, and \textit{a}$_i$ denotes the initial value of the scale factor.
The initial value of $\Omega_{\chi}$ is calculated using the widely employed shooting 
algorithm in the Boltzmann code \texttt{CLASS} \cite{1104.2932,Blas_2011}, based on 
the current value of the energy density of dark matter.
The specific deductive process can be found in reference 
\cite{Ure_a_L_pez_2016,PRD.96.061301}, for further details.
It should be noted that in the new model, the mass of SFDM is $\phi$-dependent. 
Therefore, the value of $\theta_i$ need to be adjusted accordingly.
We adopt adiabatic initial conditions for the perturbation equations of EDE and SFDM, 
for a detailed description, please refer to \cite{PRD.101.063523} and \cite{PRD.96.061301}.

\section{Numerical Results}\label{sec3}
The publicly available Boltzmann code \texttt{CLASS} \cite{1104.2932,Blas_2011} 
was modified as described in Sec.\ref{sec2}. We replace cold dark matter with 
SFDM as the constituent of dark matter. In order to compute various perturbation 
equations using the synchronous gauge in \texttt{CLASS}, we set the energy density 
fraction of cold dark matter, $\Omega_{\mathrm{cdm},0}$ to a value of $10^{-6}$ 
\cite{Ure_a_L_pez_2016}. We performed computations of the CMB power spectrum 
and the matter power spectrum by employing the existing spectrum module of 
\texttt{CLASS}.

\begin{figure}
    \includegraphics[width=\columnwidth]{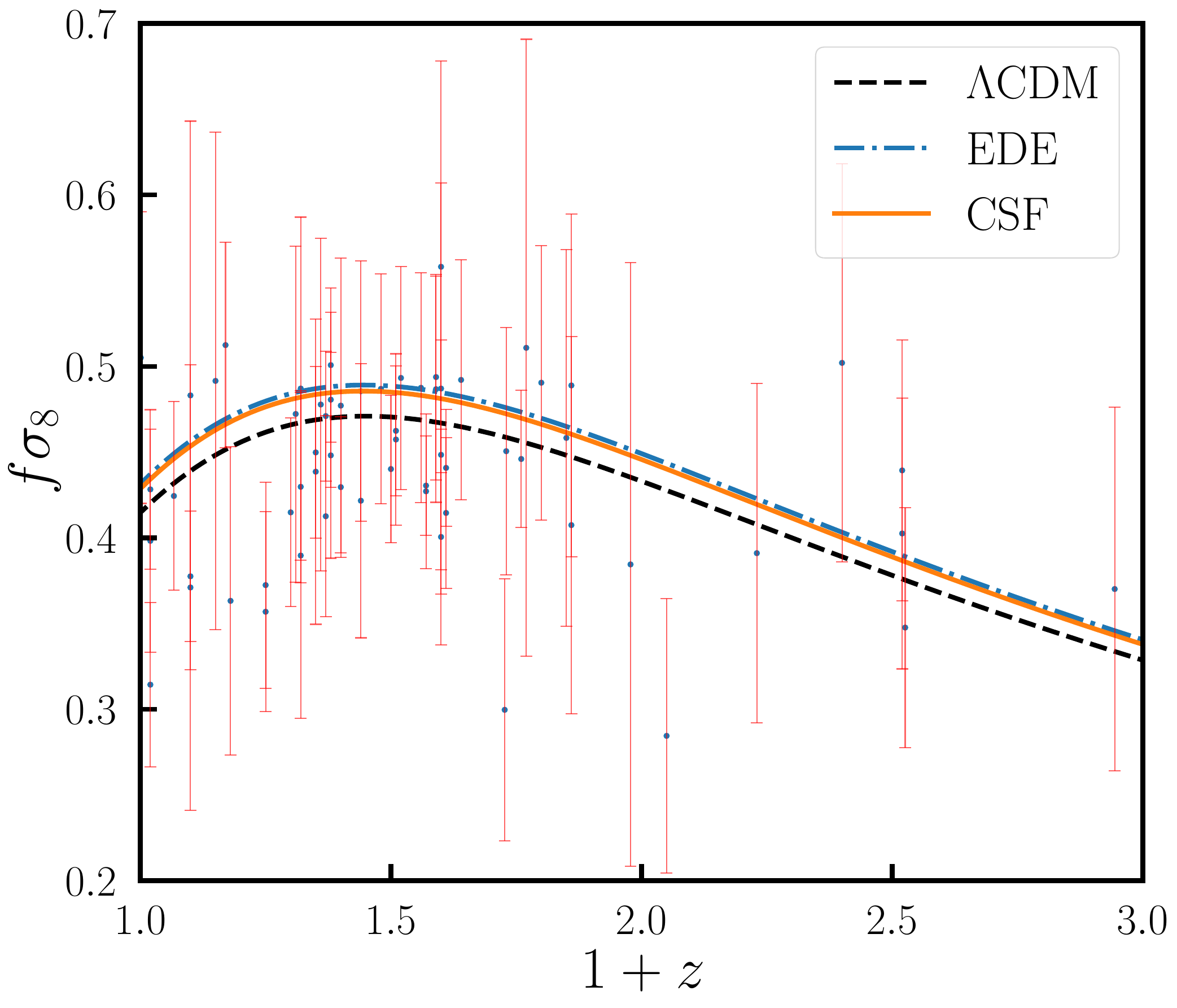}
    \caption{\label{fig4}The evolution of $f\sigma_8(z)$ with the redshift for different 
    models, namely the $\Lambda$CDM model (dashed black line), the EDE model 
    (dash-dotted blue line), and the CSF model (solid orange line). 
    }
\end{figure}

We investigate the impact of the new model on the tension of large-scale structures.
Fig.~\ref{fig4} displays the evolution of $f\sigma_8(z)$ with the redshift 
for three models, each with the corresponding best-fit values 
taken from Tab.~\ref{tab1}. The $\Lambda$CDM, EDE, and CSF models are depicted by 
dashed black, dash-dotted blue, and solid orange lines, respectively.
The 63 observed Redshift Space Distortion $f\sigma_8(z)$ data points are collected 
from \cite{PRD.97.103503}.

Compared to the $\Lambda$CDM model, both the EDE and CSF models yield larger values of 
$f\sigma_8$, exacerbating the $S_8$ tension. However, the results of the CSF model are 
slightly smaller than those of the EDE model. This discrepancy primarily stems from 
the inhibitory influence of SFDM on structure growth on small scales. We can more 
distinctly observe this characteristic in the matter power spectrum.

\begin{figure}
    \includegraphics[width=\columnwidth]{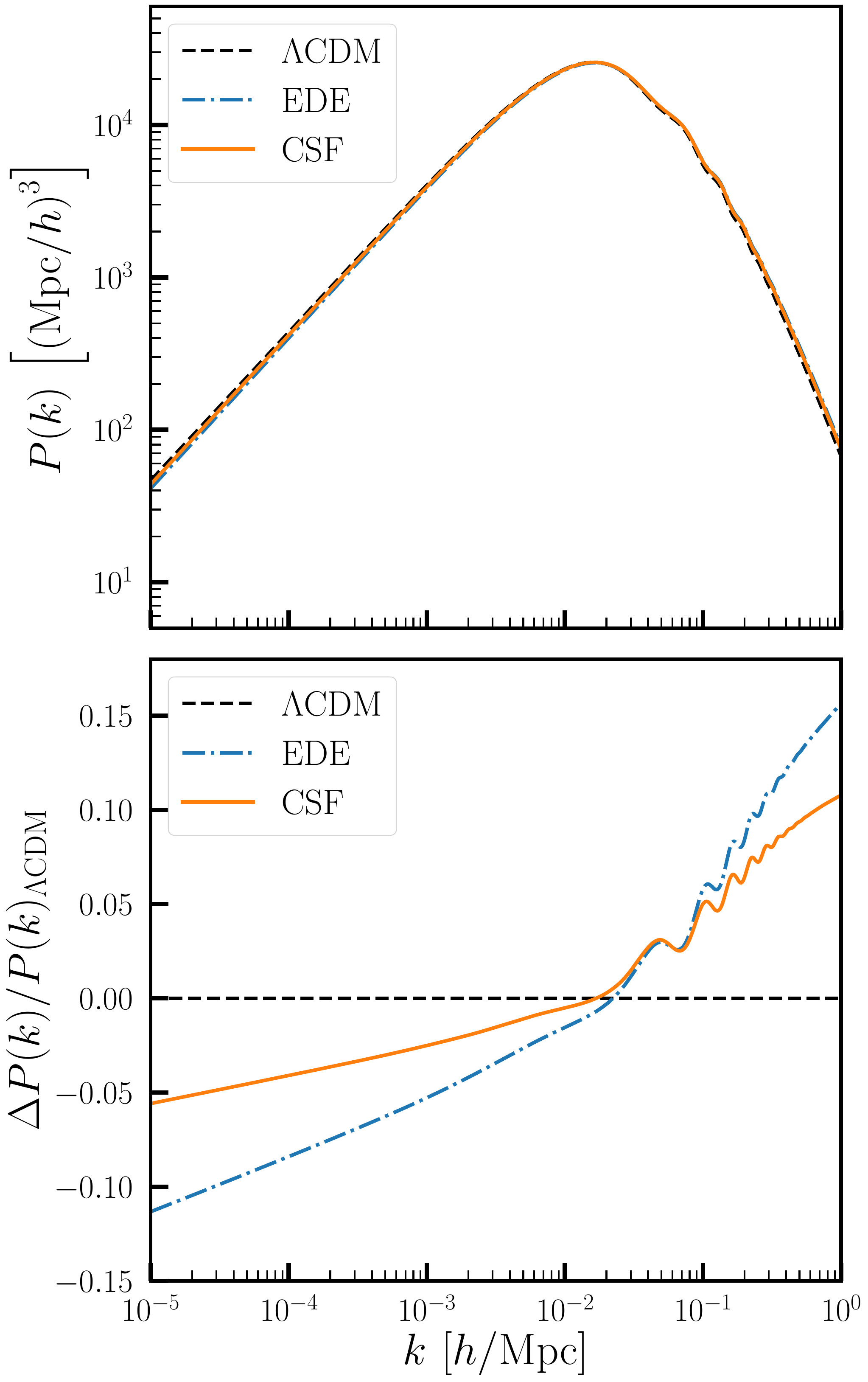}
    \caption{\label{fig5}The linear matter power spectra of three models (upper panel) 
    and their relative values compared to the $\Lambda$CDM model (lower panel) are 
    presented. The matter power spectrum obtained from the CSF 
    model is smaller than that of the EDE model on small scales. }
\end{figure}

Fig.~\ref{fig5} presents the linear matter power spectra (upper panel) and their 
relative differences compared to the $\Lambda$CDM model (lower panel) for three 
models. All parameters are obtained from the best-fit values in Tab.~\ref{tab1}. 
Due to the interplay between dark matter and dark energy, as well as 
the condensation effect of SFDM, the matter power spectrum obtained from the CSF 
model is smaller than that of the EDE model on small scales, indicating suppressed 
growth of structures and thus alleviating the $S_8$ tension caused by EDE.  
It is worth noting that the CSF model still obtains a larger power spectrum on small 
scales compared to the $\Lambda$CDM model, thus we have not resolved the  
$S_8$ tension completely.

\section{Constraints and results}\label{sec4}
The Markov Chain Monte Carlo (MCMC) 
analysis was performed using \texttt{MontePython} \cite{PRD.97.063506,Audren_2013}, 
and the MCMC chains were analyzed using \texttt{GetDist} \cite{Lewis:2019xzd}.
We conducted the analysis using the following datasets: 
\begin{itemize}
    \item[1.] \textbf{CMB}: The temperature and polarization power spectra 
    from \textit{Planck} 2018 low-$\ell$, high-$\ell$ and CMB lensing power spectrum 
    \cite{planck2020,osti_1676388,osti_1775409}.
    \item[2.] \textbf{BAO}: The measurements from BOSS-DR12 $f\sigma_8$ sample, 
    namely, the combined LOWZ and CMASS galaxy samples \cite{Alam_2017,Buen_Abad_2018} 
    and the small-z measurements from 6dFGS and the SDSS DR7 \cite{19250.x,10.1093}.
    \item[3.] \textbf{Supernovae}: The Pantheon sample, composed of 1048 supernovae Ia 
    in the redshift range 0.01 < z < 2.3 \cite{Scolnic_2018}.
\end{itemize}  

\begin{itemize}
    \item[4.] \textbf{SH0ES}: The recent SH0ES measurement with 
    $H_0 = 73.04 \pm 1.04$ km/s/Mpc \cite{Riess2021ACM}.
    \item[5.] \textbf{DES-Y3}: The $S_8 = 0.776 \pm 0.017$ from 
    Dark Energy Survey Year-3 weak lensing and galaxy clustering data \cite{PRD.105.023520}.
\end{itemize}

The results of the parameter constraints are shown in Tab.~\ref{tab1}.
\begin{table*}
    \caption{\label{tab1}The table presents the best-fit parameters and 68\% C.L. 
    marginalized constraints for $\Lambda$CDM, EDE, and CSF models using a 
    combined data comprising CMB, BAO, SNIa, SH0ES, and $S_8$ from DES-Y3. 
    The upper section of the table lists the cosmological parameters that were 
    sampled in the MCMC, while the lower section displays the derived parameters. }
    \renewcommand{\arraystretch}{1.2}
\resizebox{\linewidth}{!}{
\begin{tabular} { l  c  c  c}
    \hline
    \hline
    Model  &  $\Lambda$CDM  &  EDE  &  CSF\\
    \hline
    $100\omega{}_{b }$&
    $2.269(2.263\pm 0.014)$&
    $2.276(2.281^{+0.024}_{-0.020})$&
    $2.249(2.278^{+0.023}_{-0.018})$\\

    $\omega{}_{dm}$&
    $0.11724(0.11725\pm 0.00084)$&
    $0.1310(0.1299\pm 0.0028)$&
    $0.1285(0.1282^{+0.0024}_{-0.0028})$\\

    $H_0$&
    $68.73(68.71^{+0.35}_{-0.41})$&
    $71.85(72.46\pm 0.86)$&
    $71.13(72.20\pm 0.81)$\\

    $\ln(10^{10}A_{s})$&
    $3.043(3.050\pm 0.015)$&
    $3.057(3.063^{+0.015}_{-0.017})$&
    $3.058(3.060\pm 0.016)$\\

    $n_{s}$&
    $0.9736(0.9722\pm 0.0040)$&
    $0.9877(0.9908\pm 0.0059)$&
    $0.9804(0.9870^{+0.0067}_{-0.0050})$\\

    $\tau{}_{reio}$&
    $0.0574(0.0592\pm 0.0082)$&
    $0.0539(0.0563\pm 0.0090)$&
    $0.0554(0.0561^{+0.0082}_{-0.00035})$\\

    $\log_{10}(m_{\phi})$&
    $-$&
    $-27.292(-27.290\pm 0.055)$&
    $-27.310(-27.287\pm 0.057)$\\

    $\log_{10}(f_{\phi})$&
    $-$&
    $26.632(26.616^{+0.056}_{-0.033})$&
    $26.563(26.643\pm 0.044)$\\

    $\alpha_i $&
    $-$&
    $2.762(2.783\pm 0.069)$&
    $2.712(2.684\pm 0.053)$\\    

    $\beta$&
    $-$&
    $-$&
    $-0.027(-0.014\pm0.016)$\\

    $\log_{10}(m_{\chi})$&
    $-$&
    $-$&
    $-22.092(-22.008^{+0.087}_{-0.070})$\\  
    
    \hline

    $10^{-9}A_{s}$&
    $2.096(2.112\pm 0.032)$&
    $2.127(2.139^{+0.031}_{-0.036})$&
    $2.129(2.133\pm 0.035)$\\

    $100\theta{}_{s}$&
    $1.04206(1.04217^{+0.00025}_{-0.00031})$&
    $1.04121(1.04145\pm0.00043)$&
    $1.04138(1.04138^{+0.00029}_{-0.00035})$\\

    $f_\mathrm{EDE} $&
    $-$&
    $0.1183(0.119^{+0.023}_{-0.018})$&
    $0.1038(0.119\pm 0.022)$\\

    $\log_{10}(z_{c})$&
    $-$&
    $3.571(3.568\pm 0.034)$&
    $3.551(3.577\pm 0.034)$\\    

    $\Omega{}_{m}$&
    $0.2976(0.2977\pm 0.0048)$&
    $0.2991(0.2923\pm 0.0056)$&
    $0.2997(0.2936\pm 0.0057)$\\

    $\sigma_8$&
    $0.8017(0.8047\pm 0.0060)$&
    $0.8329(0.8325\pm 0.0083)$&
    $0.8260(0.8285^{+0.0085}_{-0.0075})$\\    

    $S_{8}$&
    $0.7985(0.8016^{+0.0096}_{-0.0080})$&
    $0.8316(0.822^{+0.011}_{-0.0093})$&
    $0.8256(0.820^{+0.014}_{-0.0080})$\\

    \hline
    $\Delta\chi^2_\mathrm{tot}$  &  $-$  &  $-11.74$  &  $-13.78$\\      
    $\Delta \mathrm{AIC}$        &  $-$  &  $ -5.74$  &  $ -3.78$\\ 
    \hline
    \hline
\end{tabular}
}
\end{table*}
The upper part of the table enlists the cosmological parameters that underwent 
sampling in the Markov chain Monte Carlo (MCMC) method. Meanwhile, the lower 
section exhibits the derived parameters. 

We use the complete dataset to ensure convergence for all models, with each parameter 
achieving the Gelman-Rubin statistic value of $R-1 < 0.05$ \cite{Gelman1992InferenceFI}.

According to the results presented in Tab.~\ref{tab1}, the EDE and CSF models obtained 
$H_0$ values of $72.46\pm 0.86$ km/s/Mpc and $72.20\pm 0.81$ km/s/Mpc at a 
68\% confidence level, respectively, which are higher than the value of 
$68.71^{+0.35}_{-0.41}$ km/s/Mpc obtained by the $\Lambda$CDM model. 
This suggests that both the EDE and CSF models can alleviate the Hubble tension. 

\begin{figure}
    \includegraphics[width=\columnwidth]{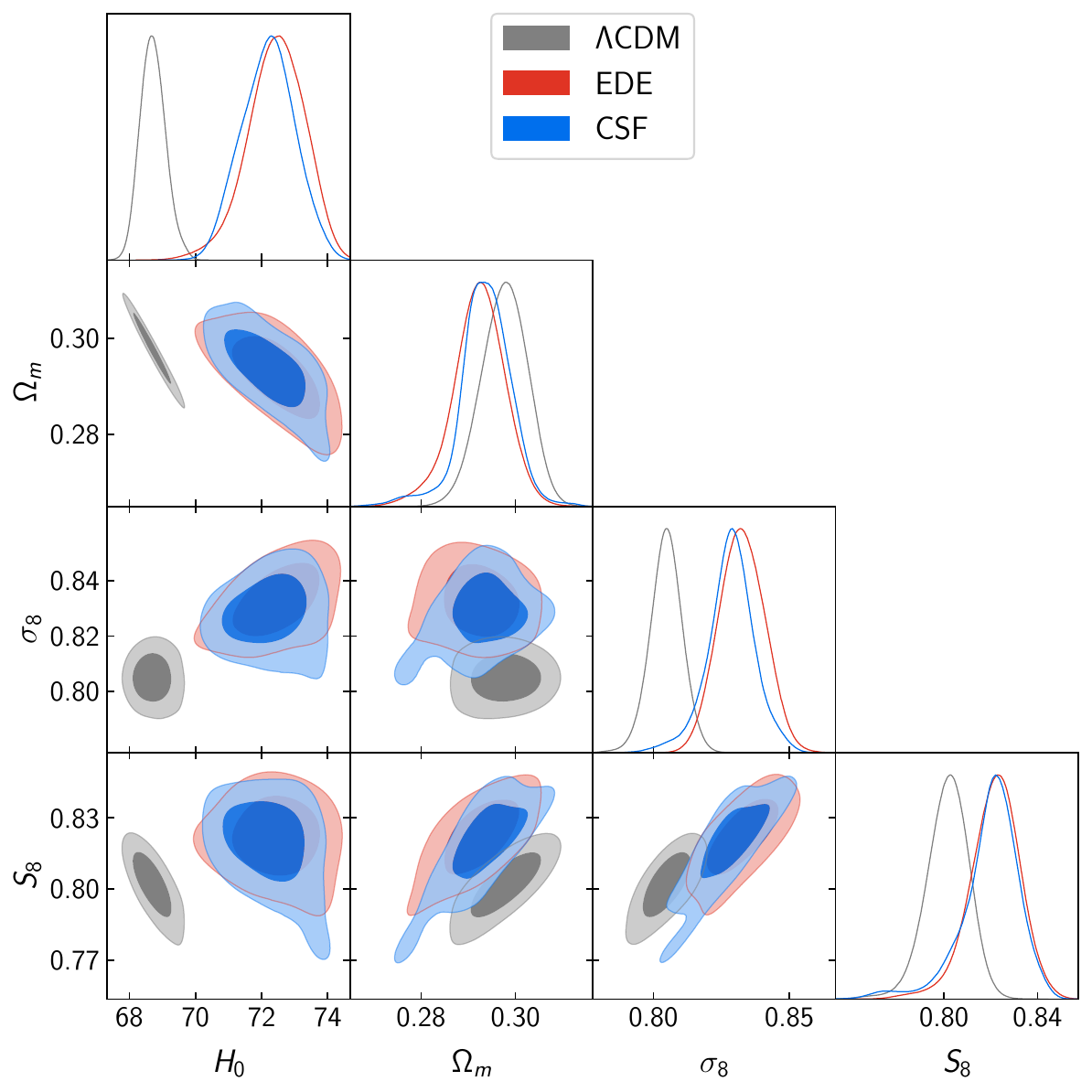}
    \caption{\label{fig6}The posteriors of various models.
    Compared to the $\Lambda$CDM model, both the EDE and CSF models yield larger $H_0$, 
    and consequently, larger $S_8$. However, the CSF model partially alleviates 
    the $S_8$ tension compared to the EDE model. 
    The full posteriors is provided in the appendix.
    }
\end{figure}

However, the EDE model and CSF model resulted in larger values of $S_8$, 
which further exacerbated the tension with the LSS. 
We obtained a non-zero coupling constant, $\beta$, with a value of 
$-0.014\pm 0.016$ at a 68\% C.L., indicating the interaction between dark 
components through the conversion of dark matter into dark energy. 
Combined with the condensation of SFDM on small scales, it is clear 
from Fig.~\ref{fig6} that the CSF model yields smaller density fluctuation amplitude 
$\sigma_8$ compared to the EDE model, thereby 
alleviating the $S_8$ tension caused by the EDE model.

The penultimate row of Tab.~\ref{tab1} displays the $\Delta \chi^2_\mathrm{tot}$ values for the EDE and CSF models 
relative to the $\Lambda$CDM model, which are $-11.74$ and $-13.78$, respectively, 
which is primarily attributed to the data from SH0ES. This 
indicates that both models fit the data better than the standard model. 
Furthermore, the $\chi^2_\mathrm{tot}$ obtained by our new model is smaller than that of the 
EDE model. This is attributed to the CSF model obtaining a smaller $S_8$ compared to 
the EDE model, resulting in a closer alignment with the data from DES-Y3. 
Thus, from a $\chi^2_\mathrm{tot}$ perspective, our novel model exhibits the performance.

We also compared the models by calculating the Akaike Information Criterion (AIC) 
\cite{Akaike_1974},
\begin{equation}
    \mathrm{AIC}=\chi^2_\mathrm{tot}+2k,
\end{equation}
where $k$ represents the number of fitted parameters. 
The smaller the AIC value of a model, the higher its goodness of fit. 
The results are presented in the last row of Tab.~\ref{tab1}. The $\Delta$AIC values for the 
EDE and CSF models relative to the $\Lambda$CDM model are $-5.74$ and $-3.78$, respectively, 
which indicates that the EDE model has the best fit.
Despite the CSF model demonstrating a smaller $\chi^2_\mathrm{tot}$ value, its performance is 
slightly inferior to that of the EDE model from the perspective of AIC, primarily 
due to the incorporation of additional parameters.

\section{Conclusion}\label{sec5}
This study examines the interplay between early dark energy (EDE) and scalar field dark 
matter (SFDM), proposing a coupled scalar fields (CSF) model to reconcile the 
discrepancies in $H_0$ and $S_8$ measurements. The CSF model leverages the EDE 
component to enhance $H_0$ without compromising the cosmic microwave background (CMB) 
observations, additionally, the suppression of SFDM on small-scale structure growth and 
the drag of dark energy on dark matter can alleviate the extra $S_8$ tension caused by EDE. 

We investigated the evolutionary equations of the coupled model, encompassing both the 
background and perturbation levels, and explored their impact on the growth of 
structures and the power spectrum of matter.
We then constrain the parameters of the $\Lambda$CDM, EDE, and CSF models using the full 
data including CMB, BAO, SNIa, SH0ES, and $S_8$ from DES-Y3. 

We constrain the coupling constant to be $-0.014\pm 0.016$ at a 68\% C.L., 
indicating the interaction between dark matter and dark energy. 
The EDE and CSF models yield $H_0$ values of $72.46\pm 0.86$ km/s/Mpc and 
$72.20\pm 0.81$ km/s/Mpc at a 68\% C.L., respectively, which are higher than the 
$\Lambda$CDM value of $68.71^{+0.35}_{-0.41}$ km/s/Mpc, thus alleviating the Hubble 
tension. 

In addition, the EDE model and CSF model yield $S_8$ best-fit values of $0.8316$ and 
$0.82 56$ respectively, both of which exceed the result of the $\Lambda$CDM model at 
$0.7985$, further exacerbating the existing $S_8$ tension. However, it is notable 
that the $S_8$ for the CSF model is lower than that of the EDE model, and the 
$\chi^2_\mathrm{tot}$ obtained from fitting the data in the former is also smaller 
than that in the latter, indicating the potential 
of the new model to alleviate the negative effect associated with the EDE model. 
We have also computed the AIC for model comparison. Despite the smaller 
$\chi^2_\mathrm{tot}$ of the CSF model, its weaker fit compared to the EDE model 
can be attributed to the introduction of additional parameters.

\begin{acknowledgments}
    This work is supported in part by National Natural Science Foundation of China 
    under Grant No.12075042, Grant No.11675032 (People's Republic of China).
\end{acknowledgments}
    
\appendix*
\section{The full MCMC posteriors}
\begin{figure*}
    \includegraphics[width=\linewidth]{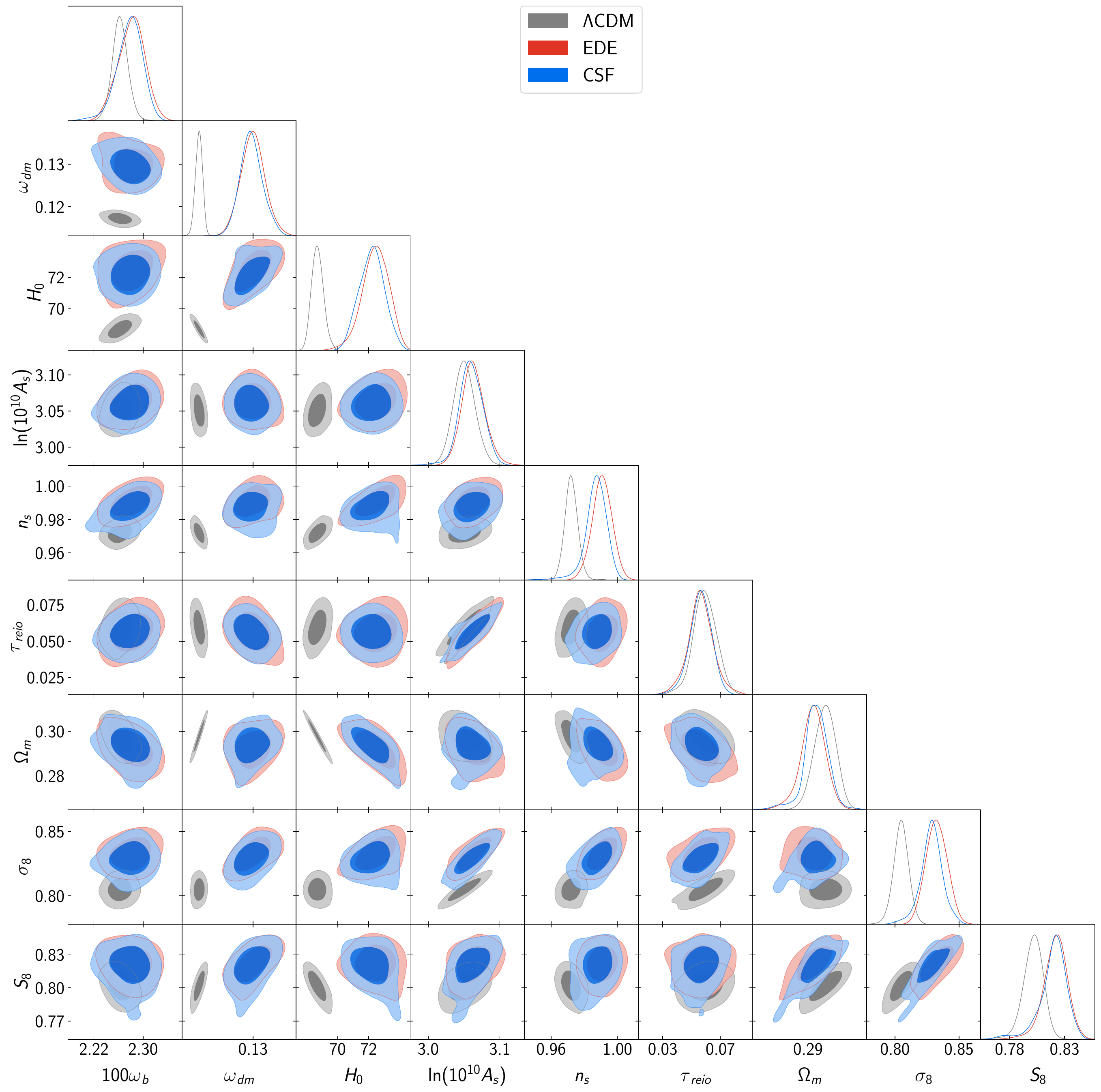}
    \caption{\label{fig7}The full Markov Chain Monte Carlo posteriors of 
    various models were obtained by utilizing the entire dataset, including CMB, BAO, 
    SNIa, SH0ES, and $S_8$ from DES-Y3.}
\end{figure*}

\bibliography{csf}

\begin{thebibliography}{61}%
\makeatletter
\providecommand \@ifxundefined [1]{%
 \@ifx{#1\undefined}
}%
\providecommand \@ifnum [1]{%
 \ifnum #1\expandafter \@firstoftwo
 \else \expandafter \@secondoftwo
 \fi
}%
\providecommand \@ifx [1]{%
 \ifx #1\expandafter \@firstoftwo
 \else \expandafter \@secondoftwo
 \fi
}%
\providecommand \natexlab [1]{#1}%
\providecommand \enquote  [1]{``#1''}%
\providecommand \bibnamefont  [1]{#1}%
\providecommand \bibfnamefont [1]{#1}%
\providecommand \citenamefont [1]{#1}%
\providecommand \href@noop [0]{\@secondoftwo}%
\providecommand \href [0]{\begingroup \@sanitize@url \@href}%
\providecommand \@href[1]{\@@startlink{#1}\@@href}%
\providecommand \@@href[1]{\endgroup#1\@@endlink}%
\providecommand \@sanitize@url [0]{\catcode `\\12\catcode `\$12\catcode `\&12\catcode `\#12\catcode `\^12\catcode `\_12\catcode `\%12\relax}%
\providecommand \@@startlink[1]{}%
\providecommand \@@endlink[0]{}%
\providecommand \url  [0]{\begingroup\@sanitize@url \@url }%
\providecommand \@url [1]{\endgroup\@href {#1}{\urlprefix }}%
\providecommand \urlprefix  [0]{URL }%
\providecommand \Eprint [0]{\href }%
\providecommand \doibase [0]{https://doi.org/}%
\providecommand \selectlanguage [0]{\@gobble}%
\providecommand \bibinfo  [0]{\@secondoftwo}%
\providecommand \bibfield  [0]{\@secondoftwo}%
\providecommand \translation [1]{[#1]}%
\providecommand \BibitemOpen [0]{}%
\providecommand \bibitemStop [0]{}%
\providecommand \bibitemNoStop [0]{.\EOS\space}%
\providecommand \EOS [0]{\spacefactor3000\relax}%
\providecommand \BibitemShut  [1]{\csname bibitem#1\endcsname}%
\let\auto@bib@innerbib\@empty
\bibitem [{\citenamefont {Verde}\ \emph {et~al.}(2019)\citenamefont {Verde}, \citenamefont {Treu},\ and\ \citenamefont {Riess}}]{Verde2019TensionsBT}%
  \BibitemOpen
  \bibfield  {author} {\bibinfo {author} {\bibfnamefont {L.}~\bibnamefont {Verde}}, \bibinfo {author} {\bibfnamefont {T.}~\bibnamefont {Treu}},\ and\ \bibinfo {author} {\bibfnamefont {A.~G.}\ \bibnamefont {Riess}},\ }\bibfield  {title} {\bibinfo {title} {Tensions between the early and late universe},\ }\href {https://doi.org/10.1038/s41550-019-0902-0} {\bibfield  {journal} {\bibinfo  {journal} {Nature Astronomy}\ } (\bibinfo {year} {2019})}\BibitemShut {NoStop}%
\bibitem [{\citenamefont {Riess}\ \emph {et~al.}(2019)\citenamefont {Riess}, \citenamefont {Casertano}, \citenamefont {Yuan} \emph {et~al.}}]{Riess_2019}%
  \BibitemOpen
  \bibfield  {author} {\bibinfo {author} {\bibfnamefont {A.~G.}\ \bibnamefont {Riess}}, \bibinfo {author} {\bibfnamefont {S.}~\bibnamefont {Casertano}}, \bibinfo {author} {\bibfnamefont {W.}~\bibnamefont {Yuan}}, \emph {et~al.},\ }\bibfield  {title} {\bibinfo {title} {Large magellanic cloud cepheid standards provide a 1\% foundation for the determination of the hubble constant and stronger evidence for physics beyond ${\Lambda}${CDM}},\ }\href {https://doi.org/10.3847/1538-4357/ab1422} {\bibfield  {journal} {\bibinfo  {journal} {The Astrophysical Journal}\ }\textbf {\bibinfo {volume} {876}},\ \bibinfo {pages} {85} (\bibinfo {year} {2019})}\BibitemShut {NoStop}%
\bibitem [{\citenamefont {Riess}\ \emph {et~al.}(2021{\natexlab{a}})\citenamefont {Riess}, \citenamefont {Casertano}, \citenamefont {Yuan} \emph {et~al.}}]{Riess_2021}%
  \BibitemOpen
  \bibfield  {author} {\bibinfo {author} {\bibfnamefont {A.~G.}\ \bibnamefont {Riess}}, \bibinfo {author} {\bibfnamefont {S.}~\bibnamefont {Casertano}}, \bibinfo {author} {\bibfnamefont {W.}~\bibnamefont {Yuan}}, \emph {et~al.},\ }\bibfield  {title} {\bibinfo {title} {Cosmic distances calibrated to 1{\%} precision with gaia {EDR}3 parallaxes and hubble space telescope photometry of 75 milky way cepheids confirm tension with ${\Lambda}${CDM}},\ }\href {https://doi.org/10.3847/2041-8213/abdbaf} {\bibfield  {journal} {\bibinfo  {journal} {The Astrophysical Journal Letters}\ }\textbf {\bibinfo {volume} {908}},\ \bibinfo {pages} {L6} (\bibinfo {year} {2021}{\natexlab{a}})}\BibitemShut {NoStop}%
\bibitem [{\citenamefont {{Planck Collaboration}}\ \emph {et~al.}(2020)\citenamefont {{Planck Collaboration}}, \citenamefont {{Aghanim, N.}}, \citenamefont {{Akrami, Y.}} \emph {et~al.}}]{planck2020}%
  \BibitemOpen
  \bibfield  {author} {\bibinfo {author} {\bibnamefont {{Planck Collaboration}}}, \bibinfo {author} {\bibnamefont {{Aghanim, N.}}}, \bibinfo {author} {\bibnamefont {{Akrami, Y.}}}, \emph {et~al.},\ }\bibfield  {title} {\bibinfo {title} {{Planck 2018 results - VI. Cosmological parameters}},\ }\href {https://doi.org/10.1051/0004-6361/201833910} {\bibfield  {journal} {\bibinfo  {journal} {Astronomy and Astrophysics}\ }\textbf {\bibinfo {volume} {641}},\ \bibinfo {pages} {A6} (\bibinfo {year} {2020})}\BibitemShut {NoStop}%
\bibitem [{\citenamefont {Cooke}\ \emph {et~al.}(2016)\citenamefont {Cooke}, \citenamefont {Pettini}, \citenamefont {Nollett},\ and\ \citenamefont {Jorgenson}}]{Cooke_2016}%
  \BibitemOpen
  \bibfield  {author} {\bibinfo {author} {\bibfnamefont {R.~J.}\ \bibnamefont {Cooke}}, \bibinfo {author} {\bibfnamefont {M.}~\bibnamefont {Pettini}}, \bibinfo {author} {\bibfnamefont {K.~M.}\ \bibnamefont {Nollett}},\ and\ \bibinfo {author} {\bibfnamefont {R.}~\bibnamefont {Jorgenson}},\ }\bibfield  {title} {\bibinfo {title} {The primordial deuterium abundance of the most metal-poor damped {L}yman-alpha system},\ }\href {https://doi.org/10.3847/0004-637x/830/2/148} {\bibfield  {journal} {\bibinfo  {journal} {The Astrophysical Journal}\ }\textbf {\bibinfo {volume} {830}},\ \bibinfo {pages} {148} (\bibinfo {year} {2016})}\BibitemShut {NoStop}%
\bibitem [{\citenamefont {Schöneberg}\ \emph {et~al.}(2019)\citenamefont {Schöneberg}, \citenamefont {Lesgourgues},\ and\ \citenamefont {Hooper}}]{Sch_neberg_2019}%
  \BibitemOpen
  \bibfield  {author} {\bibinfo {author} {\bibfnamefont {N.}~\bibnamefont {Schöneberg}}, \bibinfo {author} {\bibfnamefont {J.}~\bibnamefont {Lesgourgues}},\ and\ \bibinfo {author} {\bibfnamefont {D.~C.}\ \bibnamefont {Hooper}},\ }\bibfield  {title} {\bibinfo {title} {The {BAO}+{BBN} take on the hubble tension},\ }\href {https://doi.org/10.1088/1475-7516/2019/10/029} {\bibfield  {journal} {\bibinfo  {journal} {Journal of Cosmology and Astroparticle Physics}\ }\textbf {\bibinfo {volume} {2019}}\bibinfo  {number} { (10)},\ \bibinfo {pages} {029}}\BibitemShut {NoStop}%
\bibitem [{\citenamefont {Aubourg}\ \emph {et~al.}(2015)\citenamefont {Aubourg}, \citenamefont {Bailey}, \citenamefont {Bautista} \emph {et~al.}}]{PRD.92.123516}%
  \BibitemOpen
\bibfield  {number} {  }\bibfield  {author} {\bibinfo {author} {\bibfnamefont {E.}~\bibnamefont {Aubourg}}, \bibinfo {author} {\bibfnamefont {S.}~\bibnamefont {Bailey}}, \bibinfo {author} {\bibfnamefont {J.~E.}\ \bibnamefont {Bautista}}, \emph {et~al.} (\bibinfo {collaboration} {BOSS Collaboration}),\ }\bibfield  {title} {\bibinfo {title} {Cosmological implications of baryon acoustic oscillation measurements},\ }\href {https://doi.org/10.1103/PhysRevD.92.123516} {\bibfield  {journal} {\bibinfo  {journal} {Phys. Rev. D}\ }\textbf {\bibinfo {volume} {92}},\ \bibinfo {pages} {123516} (\bibinfo {year} {2015})}\BibitemShut {NoStop}%
\bibitem [{\citenamefont {Philcox}\ \emph {et~al.}(2020)\citenamefont {Philcox}, \citenamefont {Ivanov}, \citenamefont {Simonovi{\'{c}}},\ and\ \citenamefont {Zaldarriaga}}]{Philcox_2020}%
  \BibitemOpen
  \bibfield  {author} {\bibinfo {author} {\bibfnamefont {O.~H.}\ \bibnamefont {Philcox}}, \bibinfo {author} {\bibfnamefont {M.~M.}\ \bibnamefont {Ivanov}}, \bibinfo {author} {\bibfnamefont {M.}~\bibnamefont {Simonovi{\'{c}}}},\ and\ \bibinfo {author} {\bibfnamefont {M.}~\bibnamefont {Zaldarriaga}},\ }\bibfield  {title} {\bibinfo {title} {Combining full-shape and {BAO} analyses of galaxy power spectra: a 1.6{\%} {CMB}-independent constraint on ${H_0}$},\ }\href {https://doi.org/10.1088/1475-7516/2020/05/032} {\bibfield  {journal} {\bibinfo  {journal} {Journal of Cosmology and Astroparticle Physics}\ }\textbf {\bibinfo {volume} {2020}}\bibinfo  {number} { (05)},\ \bibinfo {pages} {032}}\BibitemShut {NoStop}%
\bibitem [{\citenamefont {Riess}\ \emph {et~al.}(2021{\natexlab{b}})\citenamefont {Riess}, \citenamefont {Yuan}, \citenamefont {Macri} \emph {et~al.}}]{Riess2021ACM}%
  \BibitemOpen
\bibfield  {number} {  }\bibfield  {author} {\bibinfo {author} {\bibfnamefont {A.~G.}\ \bibnamefont {Riess}}, \bibinfo {author} {\bibfnamefont {W.}~\bibnamefont {Yuan}}, \bibinfo {author} {\bibfnamefont {L.~M.}\ \bibnamefont {Macri}}, \emph {et~al.},\ }\bibfield  {title} {\bibinfo {title} {{A Comprehensive Measurement of the Local Value of the Hubble Constant with 1 km/s/Mpc Uncertainty from the Hubble Space Telescope and the SH0ES Team}}\ }(\bibinfo {year} {2021})\ \Eprint {https://arxiv.org/abs/2112.04510} {arXiv:2112.04510 [astro-ph.CO]} \BibitemShut {NoStop}%
\bibitem [{\citenamefont {Macaulay}\ \emph {et~al.}(2013)\citenamefont {Macaulay}, \citenamefont {Wehus},\ and\ \citenamefont {Eriksen}}]{PRL.111.161301}%
  \BibitemOpen
  \bibfield  {author} {\bibinfo {author} {\bibfnamefont {E.}~\bibnamefont {Macaulay}}, \bibinfo {author} {\bibfnamefont {I.~K.}\ \bibnamefont {Wehus}},\ and\ \bibinfo {author} {\bibfnamefont {H.~K.}\ \bibnamefont {Eriksen}},\ }\bibfield  {title} {\bibinfo {title} {Lower growth rate from recent redshift space distortion measurements than expected from planck},\ }\href {https://doi.org/10.1103/PhysRevLett.111.161301} {\bibfield  {journal} {\bibinfo  {journal} {Phys. Rev. Lett.}\ }\textbf {\bibinfo {volume} {111}},\ \bibinfo {pages} {161301} (\bibinfo {year} {2013})}\BibitemShut {NoStop}%
\bibitem [{\citenamefont {Hildebrandt}\ \emph {et~al.}(2020)\citenamefont {Hildebrandt}, \citenamefont {Köhlinger}, \citenamefont {Busch} \emph {et~al.}}]{Hildebrandt_2020}%
  \BibitemOpen
  \bibfield  {author} {\bibinfo {author} {\bibfnamefont {H.}~\bibnamefont {Hildebrandt}}, \bibinfo {author} {\bibfnamefont {F.}~\bibnamefont {Köhlinger}}, \bibinfo {author} {\bibfnamefont {J.}~\bibnamefont {Busch}}, \emph {et~al.},\ }\bibfield  {title} {\bibinfo {title} {Kids+viking-450: Cosmic shear tomography with optical and infrared data},\ }\href {https://doi.org/10.1051/0004-6361/201834878} {\bibfield  {journal} {\bibinfo  {journal} {Astronomy \& Astrophysics}\ }\textbf {\bibinfo {volume} {633}},\ \bibinfo {pages} {A69} (\bibinfo {year} {2020})}\BibitemShut {NoStop}%
\bibitem [{\citenamefont {Asgari}\ \emph {et~al.}(2020)\citenamefont {Asgari}, \citenamefont {Lin}, \citenamefont {Joachimi} \emph {et~al.}}]{KiDS}%
  \BibitemOpen
  \bibfield  {author} {\bibinfo {author} {\bibfnamefont {M.}~\bibnamefont {Asgari}}, \bibinfo {author} {\bibfnamefont {C.-A.}\ \bibnamefont {Lin}}, \bibinfo {author} {\bibfnamefont {B.}~\bibnamefont {Joachimi}}, \emph {et~al.},\ }\bibfield  {title} {\bibinfo {title} {Kids-1000 cosmology: Cosmic shear constraints and comparison between two point statistics},\ }\href {https://doi.org/10.1051/0004-6361/202039070} {\bibfield  {journal} {\bibinfo  {journal} {Astronomy \& Astrophysics}\ }\textbf {\bibinfo {volume} {645}} (\bibinfo {year} {2020})}\BibitemShut {NoStop}%
\bibitem [{\citenamefont {Hikage}\ \emph {et~al.}(2019)\citenamefont {Hikage}, \citenamefont {Oguri}, \citenamefont {Hamana} \emph {et~al.}}]{psz010}%
  \BibitemOpen
  \bibfield  {author} {\bibinfo {author} {\bibfnamefont {C.}~\bibnamefont {Hikage}}, \bibinfo {author} {\bibfnamefont {M.}~\bibnamefont {Oguri}}, \bibinfo {author} {\bibfnamefont {T.}~\bibnamefont {Hamana}}, \emph {et~al.},\ }\bibfield  {title} {\bibinfo {title} {{Cosmology from cosmic shear power spectra with Subaru Hyper Suprime-Cam first-year data}},\ }\bibfield  {journal} {\bibinfo  {journal} {Publications of the Astronomical Society of Japan}\ }\textbf {\bibinfo {volume} {71}},\ \href {https://doi.org/10.1093/pasj/psz010} {10.1093/pasj/psz010} (\bibinfo {year} {2019}),\ \bibinfo {note} {43}\BibitemShut {NoStop}%
\bibitem [{\citenamefont {Abbott}\ \emph {et~al.}(2022)\citenamefont {Abbott}, \citenamefont {Aguena}, \citenamefont {Alarcon} \emph {et~al.}}]{PRD.105.023520}%
  \BibitemOpen
  \bibfield  {author} {\bibinfo {author} {\bibfnamefont {T.~M.~C.}\ \bibnamefont {Abbott}}, \bibinfo {author} {\bibfnamefont {M.}~\bibnamefont {Aguena}}, \bibinfo {author} {\bibfnamefont {A.}~\bibnamefont {Alarcon}}, \emph {et~al.},\ }\bibfield  {title} {\bibinfo {title} {Dark energy survey year 3 results: Cosmological constraints from galaxy clustering and weak lensing},\ }\href {https://doi.org/10.1103/PhysRevD.105.023520} {\bibfield  {journal} {\bibinfo  {journal} {Phys. Rev. D}\ }\textbf {\bibinfo {volume} {105}},\ \bibinfo {pages} {023520} (\bibinfo {year} {2022})}\BibitemShut {NoStop}%
\bibitem [{\citenamefont {Valentino}\ \emph {et~al.}(2021)\citenamefont {Valentino}, \citenamefont {Mena}, \citenamefont {Pan}, \citenamefont {Visinelli}, \citenamefont {Yang}, \citenamefont {Melchiorri}, \citenamefont {Mota}, \citenamefont {Riess},\ and\ \citenamefont {Silk}}]{Di_2021}%
  \BibitemOpen
  \bibfield  {author} {\bibinfo {author} {\bibfnamefont {E.~D.}\ \bibnamefont {Valentino}}, \bibinfo {author} {\bibfnamefont {O.}~\bibnamefont {Mena}}, \bibinfo {author} {\bibfnamefont {S.}~\bibnamefont {Pan}}, \bibinfo {author} {\bibfnamefont {L.}~\bibnamefont {Visinelli}}, \bibinfo {author} {\bibfnamefont {W.}~\bibnamefont {Yang}}, \bibinfo {author} {\bibfnamefont {A.}~\bibnamefont {Melchiorri}}, \bibinfo {author} {\bibfnamefont {D.~F.}\ \bibnamefont {Mota}}, \bibinfo {author} {\bibfnamefont {A.~G.}\ \bibnamefont {Riess}},\ and\ \bibinfo {author} {\bibfnamefont {J.}~\bibnamefont {Silk}},\ }\bibfield  {title} {\bibinfo {title} {In the realm of the hubble tension—a review of solutions*},\ }\href {https://doi.org/10.1088/1361-6382/ac086d} {\bibfield  {journal} {\bibinfo  {journal} {Classical and Quantum Gravity}\ }\textbf {\bibinfo {volume} {38}},\ \bibinfo {pages} {153001} (\bibinfo {year} {2021})}\BibitemShut {NoStop}%
\bibitem [{\citenamefont {Li}\ and\ \citenamefont {Shafieloo}(2019)}]{Li_2019}%
  \BibitemOpen
  \bibfield  {author} {\bibinfo {author} {\bibfnamefont {X.}~\bibnamefont {Li}}\ and\ \bibinfo {author} {\bibfnamefont {A.}~\bibnamefont {Shafieloo}},\ }\bibfield  {title} {\bibinfo {title} {A simple phenomenological emergent dark energy model can resolve the hubble tension},\ }\href {https://doi.org/10.3847/2041-8213/ab3e09} {\bibfield  {journal} {\bibinfo  {journal} {The Astrophysical Journal Letters}\ }\textbf {\bibinfo {volume} {883}},\ \bibinfo {pages} {L3} (\bibinfo {year} {2019})}\BibitemShut {NoStop}%
\bibitem [{\citenamefont {Zhou}\ \emph {et~al.}(2022)\citenamefont {Zhou}, \citenamefont {Liu}, \citenamefont {Mu},\ and\ \citenamefont {Xu}}]{Zhouzh}%
  \BibitemOpen
  \bibfield  {author} {\bibinfo {author} {\bibfnamefont {Z.}~\bibnamefont {Zhou}}, \bibinfo {author} {\bibfnamefont {G.}~\bibnamefont {Liu}}, \bibinfo {author} {\bibfnamefont {Y.}~\bibnamefont {Mu}},\ and\ \bibinfo {author} {\bibfnamefont {L.}~\bibnamefont {Xu}},\ }\bibfield  {title} {\bibinfo {title} {{Can phantom transition at z ~ 1 restore the Cosmic concordance?}},\ }\href {https://doi.org/10.1093/mnras/stac053} {\bibfield  {journal} {\bibinfo  {journal} {Monthly Notices of the Royal Astronomical Society}\ }\textbf {\bibinfo {volume} {511}},\ \bibinfo {pages} {595} (\bibinfo {year} {2022})}\BibitemShut {NoStop}%
\bibitem [{\citenamefont {Poulin}\ \emph {et~al.}(2019)\citenamefont {Poulin}, \citenamefont {Smith}, \citenamefont {Karwal},\ and\ \citenamefont {Kamionkowski}}]{PRL.122.221301}%
  \BibitemOpen
  \bibfield  {author} {\bibinfo {author} {\bibfnamefont {V.}~\bibnamefont {Poulin}}, \bibinfo {author} {\bibfnamefont {T.~L.}\ \bibnamefont {Smith}}, \bibinfo {author} {\bibfnamefont {T.}~\bibnamefont {Karwal}},\ and\ \bibinfo {author} {\bibfnamefont {M.}~\bibnamefont {Kamionkowski}},\ }\bibfield  {title} {\bibinfo {title} {Early dark energy can resolve the hubble tension},\ }\href {https://doi.org/10.1103/PhysRevLett.122.221301} {\bibfield  {journal} {\bibinfo  {journal} {Phys. Rev. Lett.}\ }\textbf {\bibinfo {volume} {122}},\ \bibinfo {pages} {221301} (\bibinfo {year} {2019})}\BibitemShut {NoStop}%
\bibitem [{\citenamefont {Smith}\ \emph {et~al.}(2020)\citenamefont {Smith}, \citenamefont {Poulin},\ and\ \citenamefont {Amin}}]{PRD.101.063523}%
  \BibitemOpen
  \bibfield  {author} {\bibinfo {author} {\bibfnamefont {T.~L.}\ \bibnamefont {Smith}}, \bibinfo {author} {\bibfnamefont {V.}~\bibnamefont {Poulin}},\ and\ \bibinfo {author} {\bibfnamefont {M.~A.}\ \bibnamefont {Amin}},\ }\bibfield  {title} {\bibinfo {title} {Oscillating scalar fields and the hubble tension: A resolution with novel signatures},\ }\href {https://doi.org/10.1103/PhysRevD.101.063523} {\bibfield  {journal} {\bibinfo  {journal} {Phys. Rev. D}\ }\textbf {\bibinfo {volume} {101}},\ \bibinfo {pages} {063523} (\bibinfo {year} {2020})}\BibitemShut {NoStop}%
\bibitem [{\citenamefont {Hill}\ \emph {et~al.}(2020)\citenamefont {Hill}, \citenamefont {McDonough}, \citenamefont {Toomey},\ and\ \citenamefont {Alexander}}]{PRD.102.043507}%
  \BibitemOpen
  \bibfield  {author} {\bibinfo {author} {\bibfnamefont {J.~C.}\ \bibnamefont {Hill}}, \bibinfo {author} {\bibfnamefont {E.}~\bibnamefont {McDonough}}, \bibinfo {author} {\bibfnamefont {M.~W.}\ \bibnamefont {Toomey}},\ and\ \bibinfo {author} {\bibfnamefont {S.}~\bibnamefont {Alexander}},\ }\bibfield  {title} {\bibinfo {title} {Early dark energy does not restore cosmological concordance},\ }\href {https://doi.org/10.1103/PhysRevD.102.043507} {\bibfield  {journal} {\bibinfo  {journal} {Phys. Rev. D}\ }\textbf {\bibinfo {volume} {102}},\ \bibinfo {pages} {043507} (\bibinfo {year} {2020})}\BibitemShut {NoStop}%
\bibitem [{\citenamefont {Lin}\ \emph {et~al.}(2019)\citenamefont {Lin}, \citenamefont {Benevento}, \citenamefont {Hu},\ and\ \citenamefont {Raveri}}]{PRD.100.063542}%
  \BibitemOpen
  \bibfield  {author} {\bibinfo {author} {\bibfnamefont {M.-X.}\ \bibnamefont {Lin}}, \bibinfo {author} {\bibfnamefont {G.}~\bibnamefont {Benevento}}, \bibinfo {author} {\bibfnamefont {W.}~\bibnamefont {Hu}},\ and\ \bibinfo {author} {\bibfnamefont {M.}~\bibnamefont {Raveri}},\ }\bibfield  {title} {\bibinfo {title} {Acoustic dark energy: Potential conversion of the hubble tension},\ }\href {https://doi.org/10.1103/PhysRevD.100.063542} {\bibfield  {journal} {\bibinfo  {journal} {Phys. Rev. D}\ }\textbf {\bibinfo {volume} {100}},\ \bibinfo {pages} {063542} (\bibinfo {year} {2019})}\BibitemShut {NoStop}%
\bibitem [{\citenamefont {Niedermann}\ and\ \citenamefont {Sloth}(2021)}]{PhysRevD.103.L041303}%
  \BibitemOpen
  \bibfield  {author} {\bibinfo {author} {\bibfnamefont {F.}~\bibnamefont {Niedermann}}\ and\ \bibinfo {author} {\bibfnamefont {M.~S.}\ \bibnamefont {Sloth}},\ }\bibfield  {title} {\bibinfo {title} {New early dark energy},\ }\href {https://doi.org/10.1103/PhysRevD.103.L041303} {\bibfield  {journal} {\bibinfo  {journal} {Phys. Rev. D}\ }\textbf {\bibinfo {volume} {103}},\ \bibinfo {pages} {L041303} (\bibinfo {year} {2021})}\BibitemShut {NoStop}%
\bibitem [{\citenamefont {Kamionkowski}\ \emph {et~al.}(2014)\citenamefont {Kamionkowski}, \citenamefont {Pradler},\ and\ \citenamefont {Walker}}]{PRL.113.251302}%
  \BibitemOpen
  \bibfield  {author} {\bibinfo {author} {\bibfnamefont {M.}~\bibnamefont {Kamionkowski}}, \bibinfo {author} {\bibfnamefont {J.}~\bibnamefont {Pradler}},\ and\ \bibinfo {author} {\bibfnamefont {D.~G.~E.}\ \bibnamefont {Walker}},\ }\bibfield  {title} {\bibinfo {title} {Dark energy from the string axiverse},\ }\href {https://doi.org/10.1103/PhysRevLett.113.251302} {\bibfield  {journal} {\bibinfo  {journal} {Phys. Rev. Lett.}\ }\textbf {\bibinfo {volume} {113}},\ \bibinfo {pages} {251302} (\bibinfo {year} {2014})}\BibitemShut {NoStop}%
\bibitem [{\citenamefont {Marsh}(2016)}]{MARSH20161}%
  \BibitemOpen
  \bibfield  {author} {\bibinfo {author} {\bibfnamefont {D.~J.}\ \bibnamefont {Marsh}},\ }\bibfield  {title} {\bibinfo {title} {Axion cosmology},\ }\href {https://doi.org/https://doi.org/10.1016/j.physrep.2016.06.005} {\bibfield  {journal} {\bibinfo  {journal} {Physics Reports}\ }\textbf {\bibinfo {volume} {643}},\ \bibinfo {pages} {1} (\bibinfo {year} {2016})}\BibitemShut {NoStop}%
\bibitem [{\citenamefont {McDonough}\ \emph {et~al.}(2022)\citenamefont {McDonough}, \citenamefont {Lin}, \citenamefont {Hill}, \citenamefont {Hu},\ and\ \citenamefont {Zhou}}]{2112.09128}%
  \BibitemOpen
  \bibfield  {author} {\bibinfo {author} {\bibfnamefont {E.}~\bibnamefont {McDonough}}, \bibinfo {author} {\bibfnamefont {M.-X.}\ \bibnamefont {Lin}}, \bibinfo {author} {\bibfnamefont {J.~C.}\ \bibnamefont {Hill}}, \bibinfo {author} {\bibfnamefont {W.}~\bibnamefont {Hu}},\ and\ \bibinfo {author} {\bibfnamefont {S.}~\bibnamefont {Zhou}},\ }\bibfield  {title} {\bibinfo {title} {Early dark sector, the hubble tension, and the swampland},\ }\href {https://doi.org/10.1103/PhysRevD.106.043525} {\bibfield  {journal} {\bibinfo  {journal} {Phys. Rev. D}\ }\textbf {\bibinfo {volume} {106}},\ \bibinfo {pages} {043525} (\bibinfo {year} {2022})}\BibitemShut {NoStop}%
\bibitem [{\citenamefont {D'Amico}\ \emph {et~al.}(2021)\citenamefont {D'Amico}, \citenamefont {Senatore}, \citenamefont {Zhang},\ and\ \citenamefont {Zheng}}]{D_Amico_2021}%
  \BibitemOpen
  \bibfield  {author} {\bibinfo {author} {\bibfnamefont {G.}~\bibnamefont {D'Amico}}, \bibinfo {author} {\bibfnamefont {L.}~\bibnamefont {Senatore}}, \bibinfo {author} {\bibfnamefont {P.}~\bibnamefont {Zhang}},\ and\ \bibinfo {author} {\bibfnamefont {H.}~\bibnamefont {Zheng}},\ }\bibfield  {title} {\bibinfo {title} {The hubble tension in light of the full-shape analysis of large-scale structure data},\ }\href {https://doi.org/10.1088/1475-7516/2021/05/072} {\bibfield  {journal} {\bibinfo  {journal} {Journal of Cosmology and Astroparticle Physics}\ }\textbf {\bibinfo {volume} {2021}}\bibinfo  {number} { (05)},\ \bibinfo {pages} {072}}\BibitemShut {NoStop}%
\bibitem [{\citenamefont {Lucca}(2021)}]{PRD.104.083510}%
  \BibitemOpen
\bibfield  {number} {  }\bibfield  {author} {\bibinfo {author} {\bibfnamefont {M.}~\bibnamefont {Lucca}},\ }\bibfield  {title} {\bibinfo {title} {Multi-interacting dark energy and its cosmological implications},\ }\href {https://doi.org/10.1103/PhysRevD.104.083510} {\bibfield  {journal} {\bibinfo  {journal} {Phys. Rev. D}\ }\textbf {\bibinfo {volume} {104}},\ \bibinfo {pages} {083510} (\bibinfo {year} {2021})}\BibitemShut {NoStop}%
\bibitem [{\citenamefont {Beltr\'an~Jim\'enez}\ \emph {et~al.}(2021)\citenamefont {Beltr\'an~Jim\'enez}, \citenamefont {Bettoni}, \citenamefont {Figueruelo}, \citenamefont {Teppa~Pannia},\ and\ \citenamefont {Tsujikawa}}]{PRD.104.103503}%
  \BibitemOpen
  \bibfield  {author} {\bibinfo {author} {\bibfnamefont {J.}~\bibnamefont {Beltr\'an~Jim\'enez}}, \bibinfo {author} {\bibfnamefont {D.}~\bibnamefont {Bettoni}}, \bibinfo {author} {\bibfnamefont {D.}~\bibnamefont {Figueruelo}}, \bibinfo {author} {\bibfnamefont {F.~A.}\ \bibnamefont {Teppa~Pannia}},\ and\ \bibinfo {author} {\bibfnamefont {S.}~\bibnamefont {Tsujikawa}},\ }\bibfield  {title} {\bibinfo {title} {Probing elastic interactions in the dark sector and the role of ${S}_{8}$},\ }\href {https://doi.org/10.1103/PhysRevD.104.103503} {\bibfield  {journal} {\bibinfo  {journal} {Phys. Rev. D}\ }\textbf {\bibinfo {volume} {104}},\ \bibinfo {pages} {103503} (\bibinfo {year} {2021})}\BibitemShut {NoStop}%
\bibitem [{\citenamefont {T\'ellez-Tovar}\ \emph {et~al.}(2022)\citenamefont {T\'ellez-Tovar}, \citenamefont {Matos},\ and\ \citenamefont {V\'azquez}}]{2112.09337}%
  \BibitemOpen
  \bibfield  {author} {\bibinfo {author} {\bibfnamefont {L.~O.}\ \bibnamefont {T\'ellez-Tovar}}, \bibinfo {author} {\bibfnamefont {T.}~\bibnamefont {Matos}},\ and\ \bibinfo {author} {\bibfnamefont {J.~A.}\ \bibnamefont {V\'azquez}},\ }\bibfield  {title} {\bibinfo {title} {Cosmological constraints on the multiscalar field dark matter model},\ }\href {https://doi.org/10.1103/PhysRevD.106.123501} {\bibfield  {journal} {\bibinfo  {journal} {Phys. Rev. D}\ }\textbf {\bibinfo {volume} {106}},\ \bibinfo {pages} {123501} (\bibinfo {year} {2022})}\BibitemShut {NoStop}%
\bibitem [{\citenamefont {Ferreira}(2021)}]{2005.03254}%
  \BibitemOpen
  \bibfield  {author} {\bibinfo {author} {\bibfnamefont {E.~G.~M.}\ \bibnamefont {Ferreira}},\ }\bibfield  {title} {\bibinfo {title} {Ultra-light dark matter},\ }\href {https://doi.org/10.1007%2Fs00159-021-00135-6} {\bibfield  {journal} {\bibinfo  {journal} {The Astronomy and Astrophysics Review}\ }\textbf {\bibinfo {volume} {29}},\ \bibinfo {pages} {1} (\bibinfo {year} {2021})}\BibitemShut {NoStop}%
\bibitem [{\citenamefont {Vafa}(2005)}]{Vafa2005TheSL}%
  \BibitemOpen
  \bibfield  {author} {\bibinfo {author} {\bibfnamefont {C.}~\bibnamefont {Vafa}},\ }\bibfield  {title} {\bibinfo {title} {The string landscape and the swampland}\ }(\bibinfo {year} {2005})\ \Eprint {https://arxiv.org/abs/0509212} {arXiv:0509212 [hep-th]} \BibitemShut {NoStop}%
\bibitem [{\citenamefont {Palti}(2019)}]{Palti}%
  \BibitemOpen
  \bibfield  {author} {\bibinfo {author} {\bibfnamefont {E.}~\bibnamefont {Palti}},\ }\bibfield  {title} {\bibinfo {title} {The swampland: Introduction and review},\ }\href {https://doi.org/10.1002/prop.201900037} {\bibfield  {journal} {\bibinfo  {journal} {Fortschritte der Physik}\ }\textbf {\bibinfo {volume} {67}} (\bibinfo {year} {2019})}\BibitemShut {NoStop}%
\bibitem [{\citenamefont {Agrawal}\ \emph {et~al.}(2021)\citenamefont {Agrawal}, \citenamefont {Obied},\ and\ \citenamefont {Vafa}}]{PRD.103.043523}%
  \BibitemOpen
  \bibfield  {author} {\bibinfo {author} {\bibfnamefont {P.}~\bibnamefont {Agrawal}}, \bibinfo {author} {\bibfnamefont {G.}~\bibnamefont {Obied}},\ and\ \bibinfo {author} {\bibfnamefont {C.}~\bibnamefont {Vafa}},\ }\bibfield  {title} {\bibinfo {title} {${H}_{0}$ tension, swampland conjectures, and the epoch of fading dark matter},\ }\href {https://doi.org/10.1103/PhysRevD.103.043523} {\bibfield  {journal} {\bibinfo  {journal} {Phys. Rev. D}\ }\textbf {\bibinfo {volume} {103}},\ \bibinfo {pages} {043523} (\bibinfo {year} {2021})}\BibitemShut {NoStop}%
\bibitem [{\citenamefont {Aghanim}\ \emph {et~al.}(2020{\natexlab{a}})\citenamefont {Aghanim}, \citenamefont {Akrami}, \citenamefont {Ashdown} \emph {et~al.}}]{osti_1676388}%
  \BibitemOpen
  \bibfield  {author} {\bibinfo {author} {\bibfnamefont {N.}~\bibnamefont {Aghanim}}, \bibinfo {author} {\bibfnamefont {Y.}~\bibnamefont {Akrami}}, \bibinfo {author} {\bibfnamefont {M.}~\bibnamefont {Ashdown}}, \emph {et~al.},\ }\bibfield  {title} {\bibinfo {title} {Planck 2018 results. v. cmb power spectra and likelihoods},\ }\bibfield  {journal} {\bibinfo  {journal} {Astronomy and Astrophysics}\ }\textbf {\bibinfo {volume} {641}},\ \href {https://doi.org/10.1051/0004-6361/201936386} {10.1051/0004-6361/201936386} (\bibinfo {year} {2020}{\natexlab{a}})\BibitemShut {NoStop}%
\bibitem [{\citenamefont {Aghanim}\ \emph {et~al.}(2020{\natexlab{b}})\citenamefont {Aghanim}, \citenamefont {Akrami}, \citenamefont {Ashdown} \emph {et~al.}}]{osti_1775409}%
  \BibitemOpen
  \bibfield  {author} {\bibinfo {author} {\bibfnamefont {N.}~\bibnamefont {Aghanim}}, \bibinfo {author} {\bibfnamefont {Y.}~\bibnamefont {Akrami}}, \bibinfo {author} {\bibfnamefont {M.}~\bibnamefont {Ashdown}}, \emph {et~al.},\ }\bibfield  {title} {\bibinfo {title} {Planck 2018 results - viii. gravitational lensing},\ }\bibfield  {journal} {\bibinfo  {journal} {Astronomy and Astrophysics}\ }\textbf {\bibinfo {volume} {641}},\ \href {https://doi.org/10.1051/0004-6361/201833886} {10.1051/0004-6361/201833886} (\bibinfo {year} {2020}{\natexlab{b}})\BibitemShut {NoStop}%
\bibitem [{\citenamefont {Alam}\ \emph {et~al.}(2017)\citenamefont {Alam}, \citenamefont {Ata}, \citenamefont {Bailey} \emph {et~al.}}]{Alam_2017}%
  \BibitemOpen
  \bibfield  {author} {\bibinfo {author} {\bibfnamefont {S.}~\bibnamefont {Alam}}, \bibinfo {author} {\bibfnamefont {M.}~\bibnamefont {Ata}}, \bibinfo {author} {\bibfnamefont {S.}~\bibnamefont {Bailey}}, \emph {et~al.},\ }\bibfield  {title} {\bibinfo {title} {The clustering of galaxies in the completed {SDSS}-{III} baryon oscillation spectroscopic survey: cosmological analysis of the {DR}12 galaxy sample},\ }\href {https://doi.org/10.1093/mnras/stx721} {\bibfield  {journal} {\bibinfo  {journal} {Monthly Notices of the Royal Astronomical Society}\ }\textbf {\bibinfo {volume} {470}},\ \bibinfo {pages} {2617} (\bibinfo {year} {2017})}\BibitemShut {NoStop}%
\bibitem [{\citenamefont {Buen-Abad}\ \emph {et~al.}(2018)\citenamefont {Buen-Abad}, \citenamefont {Schmaltz}, \citenamefont {Lesgourgues},\ and\ \citenamefont {Brinckmann}}]{Buen_Abad_2018}%
  \BibitemOpen
  \bibfield  {author} {\bibinfo {author} {\bibfnamefont {M.~A.}\ \bibnamefont {Buen-Abad}}, \bibinfo {author} {\bibfnamefont {M.}~\bibnamefont {Schmaltz}}, \bibinfo {author} {\bibfnamefont {J.}~\bibnamefont {Lesgourgues}},\ and\ \bibinfo {author} {\bibfnamefont {T.}~\bibnamefont {Brinckmann}},\ }\bibfield  {title} {\bibinfo {title} {Interacting dark sector and precision cosmology},\ }\href {https://doi.org/10.1088/1475-7516/2018/01/008} {\bibfield  {journal} {\bibinfo  {journal} {Journal of Cosmology and Astroparticle Physics}\ }\textbf {\bibinfo {volume} {2018}}\bibinfo  {number} { (01)},\ \bibinfo {pages} {008}}\BibitemShut {NoStop}%
\bibitem [{\citenamefont {Beutler}\ \emph {et~al.}(2011)\citenamefont {Beutler}, \citenamefont {Blake}, \citenamefont {Colless} \emph {et~al.}}]{19250.x}%
  \BibitemOpen
\bibfield  {number} {  }\bibfield  {author} {\bibinfo {author} {\bibfnamefont {F.}~\bibnamefont {Beutler}}, \bibinfo {author} {\bibfnamefont {C.}~\bibnamefont {Blake}}, \bibinfo {author} {\bibfnamefont {M.}~\bibnamefont {Colless}}, \emph {et~al.},\ }\bibfield  {title} {\bibinfo {title} {The 6df galaxy survey: baryon acoustic oscillations and the local hubble constant},\ }\href {https://doi.org/10.1111/j.1365-2966.2011.19250.x} {\bibfield  {journal} {\bibinfo  {journal} {Monthly Notices of the Royal Astronomical Society}\ }\textbf {\bibinfo {volume} {416}},\ \bibinfo {pages} {3017} (\bibinfo {year} {2011})}\BibitemShut {NoStop}%
\bibitem [{\citenamefont {Ross}\ \emph {et~al.}(2015)\citenamefont {Ross}, \citenamefont {Samushia}, \citenamefont {Howlett} \emph {et~al.}}]{10.1093}%
  \BibitemOpen
  \bibfield  {author} {\bibinfo {author} {\bibfnamefont {A.~J.}\ \bibnamefont {Ross}}, \bibinfo {author} {\bibfnamefont {L.}~\bibnamefont {Samushia}}, \bibinfo {author} {\bibfnamefont {C.}~\bibnamefont {Howlett}}, \emph {et~al.},\ }\bibfield  {title} {\bibinfo {title} {The clustering of the sdss dr7 main galaxy sample-i. a 4 percent distance measure at z=0.15},\ }\href {https://doi.org/10.1093/mnras/stv154} {\bibfield  {journal} {\bibinfo  {journal} {Monthly Notices of the Royal Astronomical Society}\ }\textbf {\bibinfo {volume} {449}},\ \bibinfo {pages} {835} (\bibinfo {year} {2015})}\BibitemShut {NoStop}%
\bibitem [{\citenamefont {Scolnic}\ \emph {et~al.}(2018)\citenamefont {Scolnic}, \citenamefont {Jones}, \citenamefont {Rest}, \citenamefont {Pan} \emph {et~al.}}]{Scolnic_2018}%
  \BibitemOpen
  \bibfield  {author} {\bibinfo {author} {\bibfnamefont {D.~M.}\ \bibnamefont {Scolnic}}, \bibinfo {author} {\bibfnamefont {D.~O.}\ \bibnamefont {Jones}}, \bibinfo {author} {\bibfnamefont {A.}~\bibnamefont {Rest}}, \bibinfo {author} {\bibfnamefont {Y.~C.}\ \bibnamefont {Pan}}, \emph {et~al.},\ }\bibfield  {title} {\bibinfo {title} {The complete light-curve sample of spectroscopically confirmed {SNe} ia from pan-{STARRS}1 and cosmological constraints from the combined pantheon sample},\ }\href {https://doi.org/10.3847/1538-4357/aab9bb} {\bibfield  {journal} {\bibinfo  {journal} {The Astrophysical Journal}\ }\textbf {\bibinfo {volume} {859}},\ \bibinfo {pages} {101} (\bibinfo {year} {2018})}\BibitemShut {NoStop}%
\bibitem [{\citenamefont {Beyer}\ and\ \citenamefont {Wetterich}(2014)}]{BEYER2014418}%
  \BibitemOpen
  \bibfield  {author} {\bibinfo {author} {\bibfnamefont {J.}~\bibnamefont {Beyer}}\ and\ \bibinfo {author} {\bibfnamefont {C.}~\bibnamefont {Wetterich}},\ }\bibfield  {title} {\bibinfo {title} {Small scale structures in coupled scalar field dark matter},\ }\href {https://doi.org/https://doi.org/10.1016/j.physletb.2014.10.012} {\bibfield  {journal} {\bibinfo  {journal} {Physics Letters B}\ }\textbf {\bibinfo {volume} {738}},\ \bibinfo {pages} {418} (\bibinfo {year} {2014})}\BibitemShut {NoStop}%
\bibitem [{\citenamefont {Amendola}\ and\ \citenamefont {Barbieri}(2006)}]{AMENDOLA2006192}%
  \BibitemOpen
  \bibfield  {author} {\bibinfo {author} {\bibfnamefont {L.}~\bibnamefont {Amendola}}\ and\ \bibinfo {author} {\bibfnamefont {R.}~\bibnamefont {Barbieri}},\ }\bibfield  {title} {\bibinfo {title} {Dark matter from an ultra-light pseudo-goldsone-boson},\ }\href {https://doi.org/https://doi.org/10.1016/j.physletb.2006.08.069} {\bibfield  {journal} {\bibinfo  {journal} {Physics Letters B}\ }\textbf {\bibinfo {volume} {642}},\ \bibinfo {pages} {192} (\bibinfo {year} {2006})}\BibitemShut {NoStop}%
\bibitem [{\citenamefont {Ure{\~{n}}a-L{\'{o}}pez}\ and\ \citenamefont {Gonzalez-Morales}(2016)}]{Ure_a_L_pez_2016}%
  \BibitemOpen
  \bibfield  {author} {\bibinfo {author} {\bibfnamefont {L.~A.}\ \bibnamefont {Ure{\~{n}}a-L{\'{o}}pez}}\ and\ \bibinfo {author} {\bibfnamefont {A.~X.}\ \bibnamefont {Gonzalez-Morales}},\ }\bibfield  {title} {\bibinfo {title} {Towards accurate cosmological predictions for rapidly oscillating scalar fields as dark matter},\ }\href {https://doi.org/10.1088/1475-7516/2016/07/048} {\bibfield  {journal} {\bibinfo  {journal} {Journal of Cosmology and Astroparticle Physics}\ }\textbf {\bibinfo {volume} {2016}}\bibinfo  {number} { (07)},\ \bibinfo {pages} {048}}\BibitemShut {NoStop}%
\bibitem [{\citenamefont {Copeland}\ \emph {et~al.}(1998)\citenamefont {Copeland}, \citenamefont {Liddle},\ and\ \citenamefont {Wands}}]{PRD.57.4686}%
  \BibitemOpen
\bibfield  {number} {  }\bibfield  {author} {\bibinfo {author} {\bibfnamefont {E.~J.}\ \bibnamefont {Copeland}}, \bibinfo {author} {\bibfnamefont {A.~R.}\ \bibnamefont {Liddle}},\ and\ \bibinfo {author} {\bibfnamefont {D.}~\bibnamefont {Wands}},\ }\bibfield  {title} {\bibinfo {title} {Exponential potentials and cosmological scaling solutions},\ }\href {https://doi.org/10.1103/PhysRevD.57.4686} {\bibfield  {journal} {\bibinfo  {journal} {Phys. Rev. D}\ }\textbf {\bibinfo {volume} {57}},\ \bibinfo {pages} {4686} (\bibinfo {year} {1998})}\BibitemShut {NoStop}%
\bibitem [{\citenamefont {Ure{\~{n}}a-L{\'{o}}pez}(2016)}]{PRD.94.063532}%
  \BibitemOpen
  \bibfield  {author} {\bibinfo {author} {\bibfnamefont {L.~A.}\ \bibnamefont {Ure{\~{n}}a-L{\'{o}}pez}},\ }\bibfield  {title} {\bibinfo {title} {New perturbative method for analytical solutions in single-field models of inflation},\ }\href {https://doi.org/10.1103/PhysRevD.94.063532} {\bibfield  {journal} {\bibinfo  {journal} {Phys. Rev. D}\ }\textbf {\bibinfo {volume} {94}},\ \bibinfo {pages} {063532} (\bibinfo {year} {2016})}\BibitemShut {NoStop}%
\bibitem [{\citenamefont {Wang}\ \emph {et~al.}(2016)\citenamefont {Wang}, \citenamefont {Abdalla}, \citenamefont {Atrio-Barandela},\ and\ \citenamefont {Pavón}}]{Wang_2016}%
  \BibitemOpen
  \bibfield  {author} {\bibinfo {author} {\bibfnamefont {B.}~\bibnamefont {Wang}}, \bibinfo {author} {\bibfnamefont {E.}~\bibnamefont {Abdalla}}, \bibinfo {author} {\bibfnamefont {F.}~\bibnamefont {Atrio-Barandela}},\ and\ \bibinfo {author} {\bibfnamefont {D.}~\bibnamefont {Pavón}},\ }\bibfield  {title} {\bibinfo {title} {Dark matter and dark energy interactions: theoretical challenges, cosmological implications and observational signatures},\ }\href {https://doi.org/10.1088/0034-4885/79/9/096901} {\bibfield  {journal} {\bibinfo  {journal} {Reports on Progress in Physics}\ }\textbf {\bibinfo {volume} {79}},\ \bibinfo {pages} {096901} (\bibinfo {year} {2016})}\BibitemShut {NoStop}%
\bibitem [{\citenamefont {Koyama}\ \emph {et~al.}(2009)\citenamefont {Koyama}, \citenamefont {Maartens},\ and\ \citenamefont {Song}}]{Kazuya_Koyama_2009}%
  \BibitemOpen
  \bibfield  {author} {\bibinfo {author} {\bibfnamefont {K.}~\bibnamefont {Koyama}}, \bibinfo {author} {\bibfnamefont {R.}~\bibnamefont {Maartens}},\ and\ \bibinfo {author} {\bibfnamefont {Y.-S.}\ \bibnamefont {Song}},\ }\bibfield  {title} {\bibinfo {title} {Velocities as a probe of dark sector interactions},\ }\href {https://doi.org/10.1088/1475-7516/2009/10/017} {\bibfield  {journal} {\bibinfo  {journal} {Journal of Cosmology and Astroparticle Physics}\ }\textbf {\bibinfo {volume} {2009}}\bibinfo  {number} { (10)},\ \bibinfo {pages} {017}}\BibitemShut {NoStop}%
\bibitem [{\citenamefont {Yang}\ \emph {et~al.}(2017)\citenamefont {Yang}, \citenamefont {Pan},\ and\ \citenamefont {Mota}}]{PhysRevD.96.123508}%
  \BibitemOpen
\bibfield  {number} {  }\bibfield  {author} {\bibinfo {author} {\bibfnamefont {W.}~\bibnamefont {Yang}}, \bibinfo {author} {\bibfnamefont {S.}~\bibnamefont {Pan}},\ and\ \bibinfo {author} {\bibfnamefont {D.~F.}\ \bibnamefont {Mota}},\ }\bibfield  {title} {\bibinfo {title} {Novel approach toward the large-scale stable interacting dark-energy models and their astronomical bounds},\ }\href {https://doi.org/10.1103/PhysRevD.96.123508} {\bibfield  {journal} {\bibinfo  {journal} {Phys. Rev. D}\ }\textbf {\bibinfo {volume} {96}},\ \bibinfo {pages} {123508} (\bibinfo {year} {2017})}\BibitemShut {NoStop}%
\bibitem [{\citenamefont {Mukherjee}\ and\ \citenamefont {Banerjee}(2016)}]{Mukherjee2016InSO}%
  \BibitemOpen
  \bibfield  {author} {\bibinfo {author} {\bibfnamefont {A.}~\bibnamefont {Mukherjee}}\ and\ \bibinfo {author} {\bibfnamefont {N.}~\bibnamefont {Banerjee}},\ }\bibfield  {title} {\bibinfo {title} {In search of the dark matter dark energy interaction: a kinematic approach},\ }\href {https://doi.org/10.1088/1361-6382%2Faa54c8} {\bibfield  {journal} {\bibinfo  {journal} {Classical and Quantum Gravity}\ }\textbf {\bibinfo {volume} {34}} (\bibinfo {year} {2016})}\BibitemShut {NoStop}%
\bibitem [{\citenamefont {Olivares}\ \emph {et~al.}(2008)\citenamefont {Olivares}, \citenamefont {Atrio-Barandela},\ and\ \citenamefont {Pav\'on}}]{PhysRevD.77.063513}%
  \BibitemOpen
  \bibfield  {author} {\bibinfo {author} {\bibfnamefont {G.}~\bibnamefont {Olivares}}, \bibinfo {author} {\bibfnamefont {F.}~\bibnamefont {Atrio-Barandela}},\ and\ \bibinfo {author} {\bibfnamefont {D.}~\bibnamefont {Pav\'on}},\ }\bibfield  {title} {\bibinfo {title} {Dynamics of interacting quintessence models: Observational constraints},\ }\href {https://doi.org/10.1103/PhysRevD.77.063513} {\bibfield  {journal} {\bibinfo  {journal} {Phys. Rev. D}\ }\textbf {\bibinfo {volume} {77}},\ \bibinfo {pages} {063513} (\bibinfo {year} {2008})}\BibitemShut {NoStop}%
\bibitem [{\citenamefont {Ferreira}\ and\ \citenamefont {Joyce}(1998)}]{PRD.58.023503}%
  \BibitemOpen
  \bibfield  {author} {\bibinfo {author} {\bibfnamefont {P.~G.}\ \bibnamefont {Ferreira}}\ and\ \bibinfo {author} {\bibfnamefont {M.}~\bibnamefont {Joyce}},\ }\bibfield  {title} {\bibinfo {title} {Cosmology with a primordial scaling field},\ }\href {https://doi.org/10.1103/PhysRevD.58.023503} {\bibfield  {journal} {\bibinfo  {journal} {Phys. Rev. D}\ }\textbf {\bibinfo {volume} {58}},\ \bibinfo {pages} {023503} (\bibinfo {year} {1998})}\BibitemShut {NoStop}%
\bibitem [{\citenamefont {Hu}(1998)}]{Hu_1998}%
  \BibitemOpen
  \bibfield  {author} {\bibinfo {author} {\bibfnamefont {W.}~\bibnamefont {Hu}},\ }\bibfield  {title} {\bibinfo {title} {Structure formation with generalized dark matter},\ }\href {https://doi.org/10.1086/306274} {\bibfield  {journal} {\bibinfo  {journal} {The Astrophysical Journal}\ }\textbf {\bibinfo {volume} {506}},\ \bibinfo {pages} {485} (\bibinfo {year} {1998})}\BibitemShut {NoStop}%
\bibitem [{\citenamefont {Lesgourgues}(2011)}]{1104.2932}%
  \BibitemOpen
  \bibfield  {author} {\bibinfo {author} {\bibfnamefont {J.}~\bibnamefont {Lesgourgues}},\ }\bibfield  {title} {\bibinfo {title} {The cosmic linear anisotropy solving system (class) i: Overview}\ }(\bibinfo {year} {2011})\ \Eprint {https://arxiv.org/abs/1104.2932} {arXiv:1104.2932 [astro-ph.IM]} \BibitemShut {NoStop}%
\bibitem [{\citenamefont {Blas}\ \emph {et~al.}(2011)\citenamefont {Blas}, \citenamefont {Lesgourgues},\ and\ \citenamefont {Tram}}]{Blas_2011}%
  \BibitemOpen
  \bibfield  {author} {\bibinfo {author} {\bibfnamefont {D.}~\bibnamefont {Blas}}, \bibinfo {author} {\bibfnamefont {J.}~\bibnamefont {Lesgourgues}},\ and\ \bibinfo {author} {\bibfnamefont {T.}~\bibnamefont {Tram}},\ }\bibfield  {title} {\bibinfo {title} {The cosmic linear anisotropy solving system ({CLASS}). part {II}: Approximation schemes},\ }\href {https://doi.org/10.1088/1475-7516/2011/07/034} {\bibfield  {journal} {\bibinfo  {journal} {Journal of Cosmology and Astroparticle Physics}\ }\textbf {\bibinfo {volume} {2011}}\bibinfo  {number} { (07)},\ \bibinfo {pages} {034}}\BibitemShut {NoStop}%
\bibitem [{\citenamefont {Cede{\~{n}}o}\ \emph {et~al.}(2017)\citenamefont {Cede{\~{n}}o}, \citenamefont {Gonz{\'{a}}lez-Morales},\ and\ \citenamefont {Ure{\~{n}}a-L{\'{o}}pez}}]{PRD.96.061301}%
  \BibitemOpen
\bibfield  {number} {  }\bibfield  {author} {\bibinfo {author} {\bibfnamefont {F.~X.~L.}\ \bibnamefont {Cede{\~{n}}o}}, \bibinfo {author} {\bibfnamefont {A.~X.}\ \bibnamefont {Gonz{\'{a}}lez-Morales}},\ and\ \bibinfo {author} {\bibfnamefont {L.~A.}\ \bibnamefont {Ure{\~{n}}a-L{\'{o}}pez}},\ }\bibfield  {title} {\bibinfo {title} {Cosmological signatures of ultralight dark matter with an axionlike potential},\ }\href {https://doi.org/10.1103/PhysRevD.96.061301} {\bibfield  {journal} {\bibinfo  {journal} {Phys. Rev. D}\ }\textbf {\bibinfo {volume} {96}},\ \bibinfo {pages} {061301} (\bibinfo {year} {2017})}\BibitemShut {NoStop}%
\bibitem [{\citenamefont {Kazantzidis}\ and\ \citenamefont {Perivolaropoulos}(2018)}]{PRD.97.103503}%
  \BibitemOpen
  \bibfield  {author} {\bibinfo {author} {\bibfnamefont {L.}~\bibnamefont {Kazantzidis}}\ and\ \bibinfo {author} {\bibfnamefont {L.}~\bibnamefont {Perivolaropoulos}},\ }\bibfield  {title} {\bibinfo {title} {Evolution of the $f{\sigma}_{8}$ tension with the ${Planck}15{\Lambda}{CDM}$ determination and implications for modified gravity theories},\ }\href {https://doi.org/10.1103/PhysRevD.97.103503} {\bibfield  {journal} {\bibinfo  {journal} {Phys. Rev. D}\ }\textbf {\bibinfo {volume} {97}},\ \bibinfo {pages} {103503} (\bibinfo {year} {2018})}\BibitemShut {NoStop}%
\bibitem [{\citenamefont {St\"olzner}\ \emph {et~al.}(2018)\citenamefont {St\"olzner}, \citenamefont {Cuoco}, \citenamefont {Lesgourgues},\ and\ \citenamefont {Bilicki}}]{PRD.97.063506}%
  \BibitemOpen
  \bibfield  {author} {\bibinfo {author} {\bibfnamefont {B.}~\bibnamefont {St\"olzner}}, \bibinfo {author} {\bibfnamefont {A.}~\bibnamefont {Cuoco}}, \bibinfo {author} {\bibfnamefont {J.}~\bibnamefont {Lesgourgues}},\ and\ \bibinfo {author} {\bibfnamefont {M.}~\bibnamefont {Bilicki}},\ }\bibfield  {title} {\bibinfo {title} {Updated tomographic analysis of the integrated sachs-wolfe effect and implications for dark energy},\ }\href {https://doi.org/10.1103/PhysRevD.97.063506} {\bibfield  {journal} {\bibinfo  {journal} {Phys. Rev. D}\ }\textbf {\bibinfo {volume} {97}},\ \bibinfo {pages} {063506} (\bibinfo {year} {2018})}\BibitemShut {NoStop}%
\bibitem [{\citenamefont {Audren}\ \emph {et~al.}(2013)\citenamefont {Audren}, \citenamefont {Lesgourgues}, \citenamefont {Benabed},\ and\ \citenamefont {Prunet}}]{Audren_2013}%
  \BibitemOpen
  \bibfield  {author} {\bibinfo {author} {\bibfnamefont {B.}~\bibnamefont {Audren}}, \bibinfo {author} {\bibfnamefont {J.}~\bibnamefont {Lesgourgues}}, \bibinfo {author} {\bibfnamefont {K.}~\bibnamefont {Benabed}},\ and\ \bibinfo {author} {\bibfnamefont {S.}~\bibnamefont {Prunet}},\ }\bibfield  {title} {\bibinfo {title} {Conservative constraints on early cosmology with {MONTEPYTHON}},\ }\href {https://doi.org/10.1088/1475-7516/2013/02/001} {\bibfield  {journal} {\bibinfo  {journal} {Journal of Cosmology and Astroparticle Physics}\ }\textbf {\bibinfo {volume} {2013}}\bibinfo  {number} { (02)},\ \bibinfo {pages} {001}}\BibitemShut {NoStop}%
\bibitem [{\citenamefont {Lewis}(2019)}]{Lewis:2019xzd}%
  \BibitemOpen
\bibfield  {number} {  }\bibfield  {author} {\bibinfo {author} {\bibfnamefont {A.}~\bibnamefont {Lewis}},\ }\bibfield  {title} {\bibinfo {title} {{GetDist: a Python package for analysing Monte Carlo samples}}\ }(\bibinfo {year} {2019})\ \Eprint {https://arxiv.org/abs/1910.13970} {arXiv:1910.13970 [astro-ph.IM]} \BibitemShut {NoStop}%
\bibitem [{\citenamefont {Gelman}\ and\ \citenamefont {Rubin}(1992)}]{Gelman1992InferenceFI}%
  \BibitemOpen
  \bibfield  {author} {\bibinfo {author} {\bibfnamefont {A.}~\bibnamefont {Gelman}}\ and\ \bibinfo {author} {\bibfnamefont {D.~B.}\ \bibnamefont {Rubin}},\ }\bibfield  {title} {\bibinfo {title} {Inference from iterative simulation using multiple sequences},\ }\href {http://dx.doi.org/ 10.1214/ss/1177011136} {\bibfield  {journal} {\bibinfo  {journal} {Statistical Science}\ }\textbf {\bibinfo {volume} {7}},\ \bibinfo {pages} {457} (\bibinfo {year} {1992})}\BibitemShut {NoStop}%
\bibitem [{\citenamefont {Akaike}(1974)}]{Akaike_1974}%
  \BibitemOpen
  \bibfield  {author} {\bibinfo {author} {\bibfnamefont {H.}~\bibnamefont {Akaike}},\ }\bibfield  {title} {\bibinfo {title} {A new look at the statistical model identification},\ }\bibfield  {journal} {\bibinfo  {journal} {IEEE Transactions on Automatic Control}\ }\href {https://doi.org/10.1109/tac.1974.1100705} {10.1109/tac.1974.1100705} (\bibinfo {year} {1974})\BibitemShut {NoStop}%
\end{thebibliography}%

\end{document}